\documentclass[
 reprint,
 amsmath,
 pre,
 amssymb,
 nofootinbib,
 aps,
 longbibliography,
]{revtex4-1}

\usepackage[pdftex]{graphicx}
\usepackage{wasysym}
\usepackage{amsfonts}
\usepackage{ifsym}
\usepackage{pifont}
 \usepackage{footmisc}
\usepackage[export]{adjustbox}
\usepackage{dsfont}
\usepackage{simplewick}
\usepackage{array}
\usepackage{subcaption}
\usepackage{multirow}
\usepackage[font={small}]{caption}
\usepackage{mdframed}

\graphicspath{ {./figures/} }

\makeatletter
\newlength{\apb@width}
\newcommand{\autoparbox}[2][c]{\settowidth{\apb@width}{#2}\parbox[#1]{\apb@width}{#2}}

\makeatother

\newcommand{\namedref}[2]{\hyperref[#2]{#1~\ref*{#2}}}

\renewcommand{\Re}{\mathop{\mathrm{Re}}}
\renewcommand{\Im}{\mathop{\mathrm{Im}}}

\newcommand{\Csphere}{{}^\bullet\kern-1.2pt C}
\newcommand{\Ctorus}{{}^\circ\kern-1.2pt C}

\newcommand{\nn}{\nonumber}

\newcommand{\COMMENT}[1]{}

\newcommand{\neqa}{\nonumber\end{eqnarray}}

\newcommand{\<}{{\langle}}
\renewcommand{\>}{{\rangle}}

\newcommand{\cA}{{\cal A}}

\newcommand{\re}{\relax{\rm I\kern-.18em R}}

\def\su2{{SU(2)}}

\def\[{\left[}
\def\]{\right]}

\def\({\left(}
\def\){\right)}
\def\[{\left[}
\def\]{\right]}

\def\<{\langle}
\def\>{\rangle}

\def\i2{\frac{i}{2}}

\def\cT{{\cal T }}

\def\2F1{\,_2{\rm F}_1}

\usepackage{ulem}

\usepackage[usenames,dvipsnames]{xcolor}
\usepackage{color}
\usepackage{graphicx}
\usepackage{dcolumn}
\usepackage{bm}
\usepackage{hyperref}

\usepackage{cancel}
\usepackage{ctable}
\usepackage{booktabs}
\newcolumntype{L}[1]{>{\raggedright\let\newline\\\arraybackslash\hspace{0pt}}m{#1}}
\newcolumntype{C}[1]{>{\centering\let\newline\\\arraybackslash\hspace{0pt}}m{#1}}
\newcolumntype{R}[1]{>{\raggedleft\let\newline\\\arraybackslash\hspace{0pt}}m{#1}}

\newcommand{\beq}{\begin{equation}}
\newcommand{\eeq}{\end{equation}}
\newcommand{\beqq}{\begin{equation*}}
\newcommand{\eeqq}{\end{equation*}}
\newcommand\beqa{\begin{eqnarray}}
\newcommand\eeqa{\end{eqnarray}}
\newcommand\beqaa{\begin{eqnarray*}}
\newcommand\eeqaa{\end{eqnarray*}}
\newcommand\bea{\begin{array}}
\newcommand\eea{\end{array}}

\usepackage{MorrisIn,lettrine,Romantik}

\begin{document}

\begin{flushleft}
	\hfill \\
    \hfill \parbox[c]{40mm}{CERN-TH-2024-177} \\ 
    \hfill \parbox[c]{40mm}{CALT-TH-2024-038}
\end{flushleft}

\title{From data to the analytic S-matrix:\\
A Bootstrap fit of the pion scattering amplitude}

\author{Andrea Guerrieri$^{a,b,c,d}$, Kelian H\"aring$^{a,e,f}$, and Ning Su$^{g,h}$}

\affiliation{
$^a$CERN, Theoretical Physics Department, CH-1211 Geneva 23, Switzerland\\
$^b$ Department of Mathematics, City, University of London, Northampton Square, EC1V
0HB, London, UK\\
$^c$Perimeter Institute for Theoretical Physics, 31 Caroline St N Waterloo, Ontario N2L 2Y5, Canada\\
$^d$Dipartimento di Fisica e Astronomia, Universita degli Studi di Padova, Italy\\
$^e$Institute for Theoretical Physics, University of Amsterdam, 
1090 GL Amsterdam, The Netherlands\\
$^{f}$Fields and Strings Laboratory, Institute of Physics, École Polytechnique Fédéral de Lausanne (EPFL), CH-1015 Lausanne, Switzerland\\
$^g$ Walter Burke Institute for Theoretical Physics, Caltech, Pasadena, California 91125, USA\\
$^h$ Department of Physics, Massachusetts Institute of Technology, Cambridge, MA 02139, USA
}


\begin{abstract}
We propose a novel strategy to fit experimental data using a UV complete amplitude ansatz satisfying the constraints of Analyticity, Crossing, and Unitarity. We focus on $\pi\pi$ scattering combining both experimental and lattice data. The fit strategy requires using S-matrix Bootstrap methods and non-convex Particle Swarm Optimization techniques. Using this procedure, we numerically construct a full-fledged scattering amplitude that fits the data and contains the known QCD spectrum that couples to $\pi \pi$ states below $1.4$ GeV. 
The amplitude constructed agrees below the two-particle threshold with the two-loop $\chi$PT prediction. Moreover, we correctly predict the $D_2$ phase shift, the appearance of a spin three state, and the behavior of the high-energy total cross-section. Finally, we find a genuine tetraquark resonance around 2 GeV, which we argue might be detected by looking into the decays of B mesons. 
\end{abstract}

\pacs{Valid PACS appear here}
\maketitle

\section{Introduction}

\lettrine{S}{cattering} data are measured in experiments for real energies. 
Yet, the physics is often hidden deep in the complex energy plane intertwined with the analytic structure of scattering amplitudes. 
The S-matrix 
program provides the tools to analytically continue measurements, and learn about the spectrum and interactions of a theory.

In this Letter, we discuss a novel approach to fit experimental scattering data using a UV complete ansatz that embodies the physical constraints of Analyticity, Crossing symmetry, and Unitarity.\footnote{To the best of our knowledge, a fit ansatz satisfying all these constraints has never been constructed. The closest representation satisfying some of these properties is in terms of Roy equations \cite{Roy:1971tc,Ananthanarayan:2000ht}.}
We propose using non-perturbative S-matrix Bootstrap methods to construct a fit model that satisfies all those properties, with only a few tunable parameters. Those parameters are then fixed by minimizing the $\chi^2$ with the experimental data. We will call this strategy ``Bootstrap Fit''. 

We test our algorithm on $\pi \pi$ scattering data.
Pions are pseudo-Goldstone bosons of the original chiral symmetry of the QCD Lagrangian. 
For simplicity, we assume the pions have all mass $m_\pi=1$,
neglecting the splitting due to isospin-breaking effects. 
Thus, they are in the vector representation of $O(3)$ and the $2\to 2$ amplitude of $\pi^a\pi^b\to \pi^c\pi^d$ is 
\begin{equation}
    \cT_{ab}^{cd}(s,t,u) {=} \cA(s|t,u)\delta_{ab}\delta^{cd} + \cA(t|s,u)\delta_a^c\delta_b^d + \cA(u|s,t)\delta_a^d\delta_b^c\,
\label{pipi_amplitude}
\end{equation}
where $s,t,u$ are the Mandelstam variables and $s{+}t{+}u{=}4$. By crossing symmetry, we also have $\cA(s|t,u)=\cA(s|u,t)$.

  \begin{figure}[t]
    \centering
    \includegraphics[width=\linewidth]{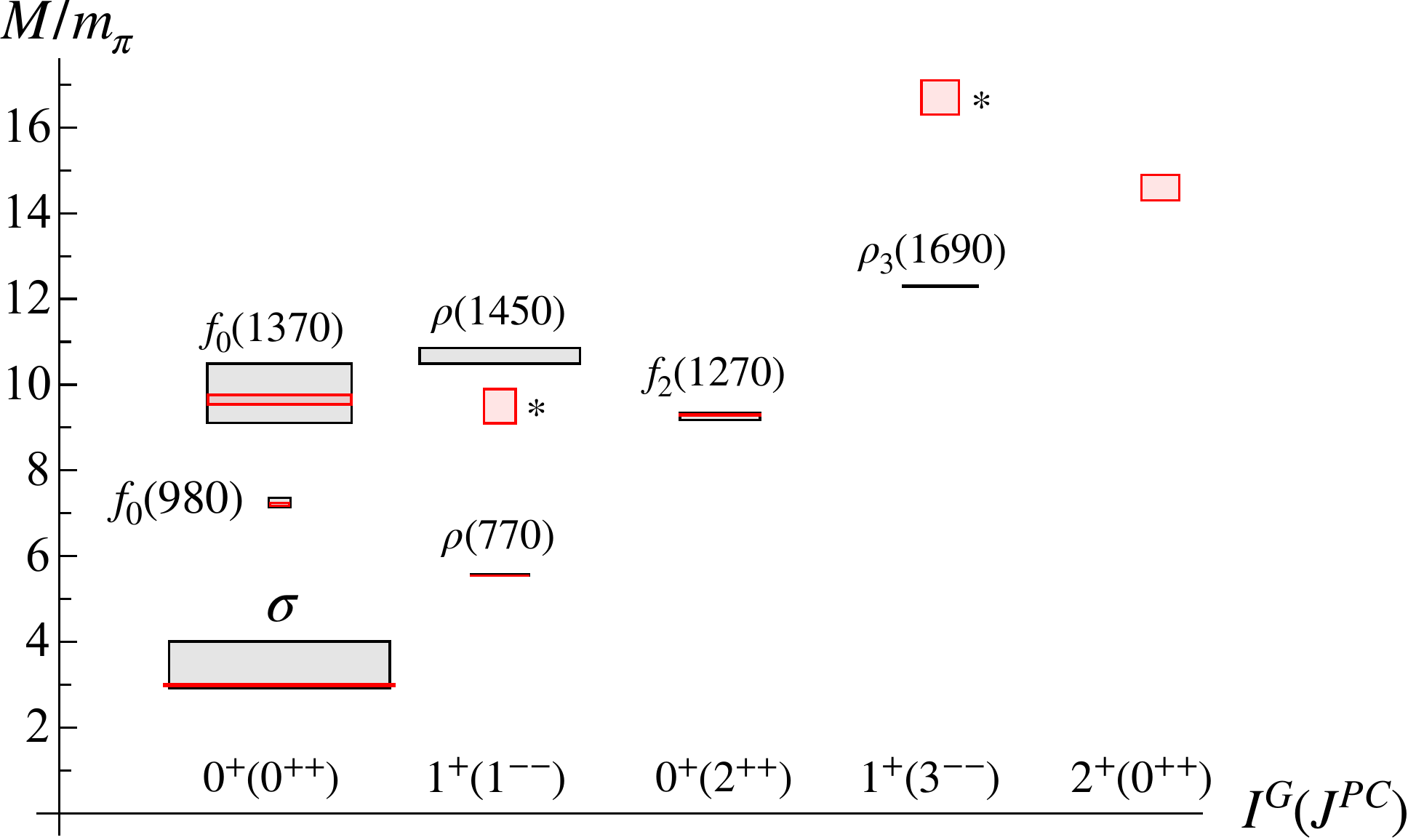}
    \caption{
    The vertical side of the boxes corresponds to the uncertainty of the mass determination, the horizontal side to $\Gamma/2$, with $\Gamma$ the width (we do not show the error on the width here). Mass and width are in the same units. \textcolor{gray}{Gray} refers to the experimental spectrum, \textcolor{red}{red} to our Bootstrap estimate. The particles marked by an asterisk, are still affected by numerical systematic as explained later in section \ref{sec:predictions}.}   
    \label{fig:spectrum}
\end{figure}

Pions are the lightest states of the QCD spectrum. A two-particle $\pi\pi$ state has $G$-parity $G=+1$, and parity $P=(-1)^J$, where $J$ is the spin of the particle. 
Only resonances with the same quantum numbers can be produced in a scattering experiment. In figure \ref{fig:spectrum}, the experimentally observed spectrum below $1.4$ GeV is represented with gray boxes, adding only the $\rho_3$ above this energy.\footnote{To express the spectrum in dimensionless units we use $m_\pi=137.3$ MeV, the average between charged and neutral pions.} 

The Numerical Bootstrap of pion scattering amplitudes was recently revived in \cite{Andrea} and extended to the massless case in \cite{Guerrieri:2020bto}.
In \cite{Bose:2020shm}, the authors refined the allowed region found in \cite{Andrea} using additional positivity constraints at low energy. By performing hypothesis testing using relative entropy, they selected the region of low energy parameters that matches QCD low energy data and $\chi$PT. In \cite{Bose:2020cod}, they have also extracted the spectrum finding hints of emerging Regge trajectories. In \cite{He:2023lyy, He:2024nwd}, the authors used several low energy constraints below the two-particle threshold, and form factors constraints computed in perturbative QCD at energy above 2 GeV, to select a region in parameter space that nicely agrees with experimental phase shifts and low energy data. The Bootstrap of large-$N$ $\pi\pi$ amplitudes was kicked off in \cite{Albert:2022oes} and \cite{Fernandez:2022kzi}, and generalized by including photons and matching with the chiral anomaly in \cite{Albert:2023jtd}. Finally, including a minimal amount of spectrum assumptions, a Bootstrap candidate for the large-$N$ QCD amplitude was found in \cite{Albert:2023seb}.

\begin{figure}[t]
    \centering   \includegraphics[width=\linewidth]{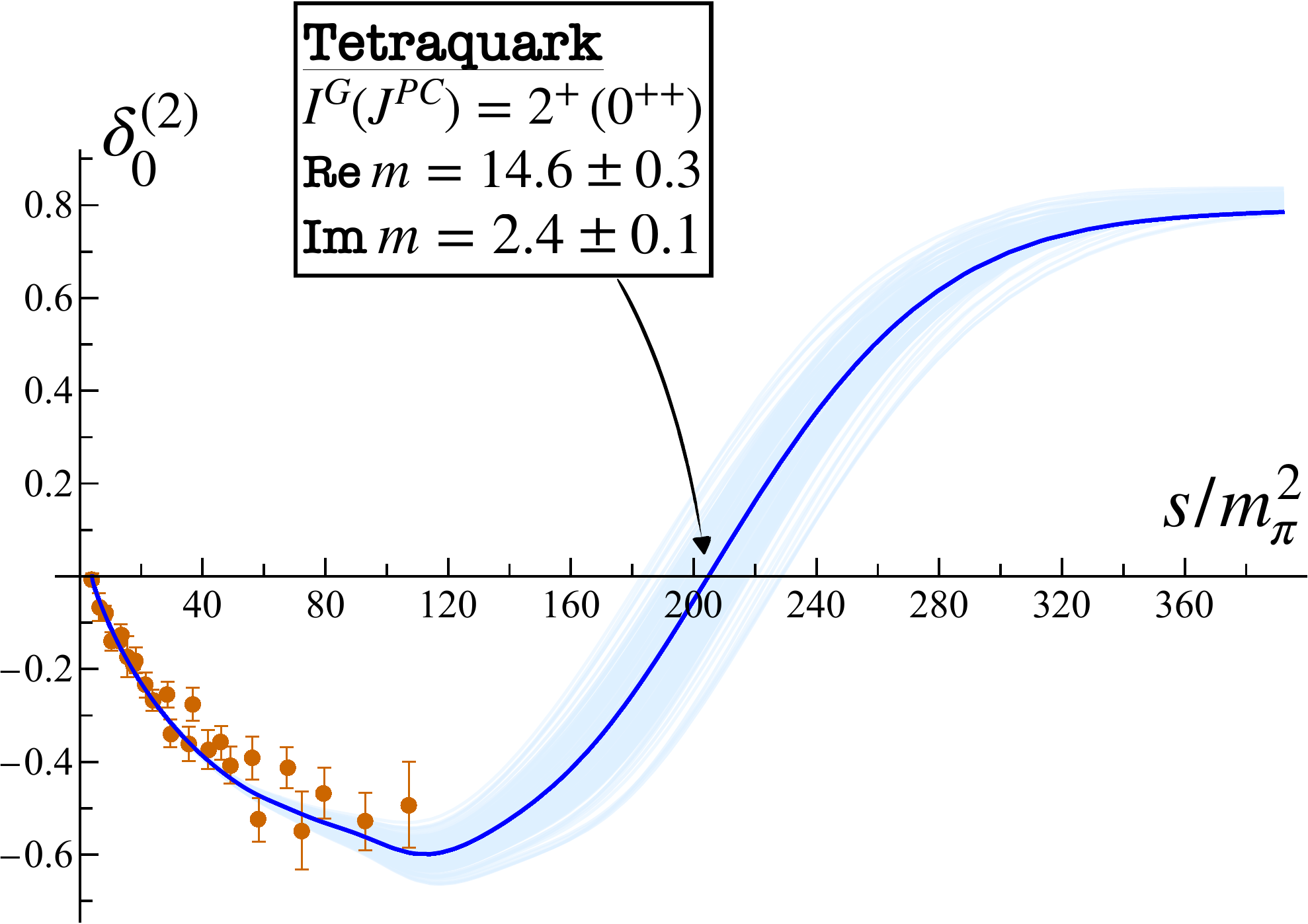}
    \caption{High energy $S_2$ partial wave phase shift profile. In blue the best fit, in light blue all amplitudes with suboptimal $\chi^2$ (see section \ref{sec:best_fit}). We see a jump in the phase around $s\approx 200$ that signals the presence of a resonance. In the inset, the mass parameters are extracted from $m=\sqrt{s^*}$, where $s^*$ is the position of the zero $S_0^{(2)}(s^*)=0$ in the complex $s$-plane, and in units of $m_\pi=1$.}
    \label{fig:Tetra_Phase}
\end{figure}

Our main result is the construction of a full-fledged analytic amplitude $\cT_{ab}^{cd}(s,t,u)$ that fits the experimental data. From this amplitude, we predict the spectrum and the low energy parameters. In figure \ref{fig:spectrum}, we present in red the Bootstrap estimate of the spectrum. Both the mass and the width of the resonances match the experimental measurements, except for those denoted by an asterisk still affected by numerical systematics. The last column has the exotic quantum numbers $2^+(0^{++})$. There is no such state in the PDG \cite{ParticleDataGroup:2024cfk}.
Because of the $I=2$ quantum number, this particle is a genuine \emph{Tetraquark} state \cite{Jaffe:1976ig,tHooft:2008rus} with a mass of around 2 GeV and a width of 600 MeV.
As a prelude to our results, in figure \ref{fig:Tetra_Phase} we show the predicted phase shift in the $S_2$ wave.
At first glance, it would be difficult to anticipate from the experimental data that the phase shift would reverse direction at higher energies.
This behavior is a genuine prediction of our Bootstrap procedure which incorporates full crossing and unitarity of the scattering amplitude. 
Before discussing in detail the physics of this amplitude we shall describe the steps of our strategy. 

In section \ref{sec:ansatz}, we discuss the S-matrix Bootstrap step in constructing a candidate amplitude $\mathcal{A}^\text{ansatz}_\Theta(s|t,u)$ depending on a small set of free parameters $\Theta$ by solving a semi-definite optimization problem. 
In section \ref{sec:PSO}, we describe the $\chi^2(\Theta)$ minimization procedure
using a gradient-free Particle Swarm Optimization (PSO) algorithm.
To evaluate the $\chi^2(\Theta)$ function we use the result of the Bootstrap problem $\mathcal{A}^\text{ansatz}_\Theta(s|t,u)$, and the experimental and lattice data available for the $S_0$, $S_2$, $P$, and $D_0$ waves, as explained in section \ref{sec:best_fit}, where we present the result of the Bootstrap Fit. In section \ref{sec:predictions} we test the amplitude obtained in different regimes by comparing it with Chiral Perturbation Theory ($\chi$PT) and the experimental measurements not used in the fitting procedure, always finding a good qualitative agreement. Finally, in section \ref{sec:discussion} we comment on possible improvements and future directions.

\section{Building the fit model}
\label{sec:ansatz}

\subsection{The analytic Bootstrap ansatz}

We parametrize the pion amplitude with the $\rho$-ansatz \cite{Paper3,Andrea}  constructed to be analytic and crossing symmetric.\footnote{Here we assume maximal analyticity. See \cite{Guerrieri:2021tak,EliasMiro:2023fqi,Guerrieri:2023qbg} for a complementary Bootstrap approach that uses only the rigorous analyticity domain proven by Martin \cite{Martin2}.}
For our problem, it is convenient to introduce an extended version of the $\rho$-ansatz, the \emph{multi-foliation} ansatz.
We call \emph{foliation}, a sum of the form
\beqa
&&\mathcal{F}^N_\sigma(s|t,u){=}\sum_{0\leq n{+}m\leq N}\alpha^{(\sigma)}_{n,m}\rho_\sigma(s)^n(\rho_\sigma(t)^m{+}\rho_\sigma(u)^m) +\nn \\
&&\sum_{\substack{n+m\leq N \\ 1\leq n\leq m}}\beta^{(\sigma)}_{n,m}(\rho_\sigma(t)^n \rho_\sigma(u)^m{+}\rho_\sigma(u)^n\rho_\sigma(t)^m),
\eeqa
where $\rho_\sigma(s)=\tfrac{\sqrt{\sigma-4}-\sqrt{4-s}}{\sqrt{\sigma-4}+\sqrt{4-s}}$ 
is a conformal map from the cut plane to the unit disk with center $\rho_\sigma(8{-}\sigma)=0$. 
$\mathcal{F}^N_\sigma(s|t,u)$ is manifestly symmetric in $t,u$ which automatically enforces crossing symmetry.
The multi-foliation ansatz is then obtained by summing over different centers $\sigma\in \Sigma$
\beq
\mathcal{A}^\text{ansatz}(s|t,u)=\sum_{\sigma\in \Sigma} \mathcal{F}^{N_\sigma}_\sigma(s|t,u).
\label{multi-ansatz}
\eeq
The intuition behind this operation is simple. A single foliation $\mathcal{F}^N_\sigma$ approximates best the amplitude in the region $|s|\approx \sigma$. For single-scale problems, tuning $\sigma$ to the desired scale is enough to achieve fast convergence \cite{EliasMiro:2022xaa,Gaikwad:2023hof}.\footnote{By fast convergence we mean that there is a $\sigma$ which minimizes the difference between the true amplitude and the foliation $|\mathcal{A}-\mathcal{F}^N_\sigma|$ at fixed $N$.}

The pion amplitude is a multi-scale function: it features chiral physics at the scale $s\approx 1$, several sharp resonances of different spin at $s \propto \Lambda^2_{QCD}$ between $s\approx 30$ and $s\approx 100$, and inelastic effects kicking in at the $K\bar K$ threshold $s\approx 50$.\footnote{It is also possible to consider foliations with different branch points. This might help to encapsulate the $K\bar K$ threshold, and obtain a better fit of the data (see also section  \ref{sec:best_fit}).} Therefore, we choose $\Sigma=\{20/3,30,50,86\}$.  
We detail the numerical implementation in Appendix \ref{sec:numerical_details}.

\subsection{The unitarity constraints}

The ansatz \eqref{multi-ansatz} is not manifestly unitary for arbitrary values of the free parameters $\alpha^{(\sigma)}_{n,m}$ and $\beta^{(\sigma)}_{n,m}$. 
We impose unitarity as a numerical constraint on those coefficients.
This is conveniently performed by projecting the amplitude on a set of partial waves diagonalizing the amplitude in angular momentum and flavor.
First, we decompose the amplitude into the isospin channels
\beqa
\mathcal{T}^{(0)}(s,t,u)&=&3\mathcal{A}(s|t,u)+\mathcal{A}(t|s,u)+\mathcal{A}(u|s,t),\\
\mathcal{T}^{(1)}(s,t,u)&=&\mathcal{A}(t|s,u)-\mathcal{A}(u|s,t),\\
\mathcal{T}^{(2)}(s,t,u)&=&\mathcal{A}(t|s,u)+\mathcal{A}(u|s,t),
\eeqa
and then project into partial waves
\begin{equation}
t_\ell^{(I)}(s)=\frac{1}{32\pi}\int_{-1}^1 d\cos\theta P_\ell(\cos\theta)\mathcal{T}^{(I)}(s,\theta)\,, 
\label{eq:parital_wave_projection}
\end{equation}
where $\cos\theta=1+2t/(s-4)$.
Finally, unitarity for partial waves becomes the probability conservation condition $|S_\ell^{(I)}|\leq 1$ for any $s>4$, any $\ell$, and $I$, with $S_\ell^{(I)}=1+i \sqrt{\tfrac{s-4}{s}}t_\ell^{(I)}$.

Experimental data show a pronounced inelasticity around $s\approx 50$ at the $K\bar K$ threshold where the process $\pi\pi \to K\bar K$ goes on-shell,
especially in the $S_0$ channel. The probability conservation must be then replaced by the stronger condition $|S_\ell^{(I)}|\leq \eta_\ell^{(I)}$. We impose this condition as an SDP inequality of the form\footnote{The effect of inelastic constraints in the S-matrix Bootstrap has been discussed in \cite{Antunes:2023irg}. The Bootstrap of full multi-particle processes has been only recently developed for branon scattering amplitudes in two space-time dimensions \cite{Guerrieri:2024ckc}. Inelasticities were also obtained as an output of a Bootstrap procedure in \cite{Tourkine:2023xtu}. }
\beq
\mathcal{U}_\ell^{(I)}=\begin{pmatrix}
\eta_\ell(s) + \Im S(s)_\ell & \Re S(s)_\ell \\
\Re S(s)_\ell & \eta_\ell(s)- \Im S(s)_\ell
\end{pmatrix}\succeq 0.
\label{eq:unitarity}
\eeq
As for the functions $\eta_\ell^{(I)}(s)$, we use the mean value of the phenomenological parametrization for the ``big-dip'' scenario discussed in \cite{Garcia-Martin:2011iqs}.\footnote{For simplicity, we do not take into account the error on this parametrization.
} These functions should be regarded as an additional phenomenological input.\footnote{It would be interesting to consider the general mixed system of pions and kaons, and obtain the inelasticity as a result of the procedure. See \cite{Homrich:2019cbt,Bercini:2019vme,Guerrieri:2020kcs} for S-matrix Bootstrap works on mixed-amplitudes.}

\subsection{Soft theorems and Spectrum Assumptions}
\label{sec:soft_and_spectrum}

In \cite{Andrea}, it was shown that the low energy parameters of the pion amplitude, such as scattering lengths, are well inside the allowed region determined by the general S-matrix Bootstrap constraints. To restrict the allowed region, it is necessary to impose additional conditions on the amplitude. To this end, we consider two types of constraints: \emph{soft theorems}, and \emph{spectrum assumptions}.

Soft theorems are the consequence of spontaneously broken chiral symmetry. If pions were massless, the amplitude would vanish $\mathcal{A}\to 0$ when any of the momenta of the particles become soft. 
As the pions are massive, this behavior is corrected by quantum effects and is no longer exact \cite{Adler:1964um}. Nevertheless, the existence of low energy zeroes in the partial waves is a prediction of $\chi$PT
\cite{Weinberg:1978kz}.
For this reason, we impose that $t_0^{(0)}(z_0)=0$, and $t_0^{(2)}(z_2)=0$ for some $0\leq z_0, z_2 \leq 4$. We refer to these constraints as \emph{chiral zeroes} conditions.\footnote{At tree-level in $\chi$PT: $z_0=1/2$, and $z_2=2$.}

The physical spectrum is encoded into the position of pole singularities in the second sheet of the $2\to 2$ scattering amplitude. 
Using the elastic unitarity condition, it is possible to relate those resonance poles to zeros of the $S$-matrix (not the amplitude!) in the first sheet.
We impose several conditions of the form $S_\ell^{(I)}(s_R)=0$, that we call \emph{resonance zeros}, where $s_R$ is the complex mass squared of the particle, and $(\ell, I)$ its quantum numbers.

At this point, it is important to stress that we are agnostic about the values of both the chiral and resonance zeros. In our construction, we only assume their existence with the corresponding quantum numbers. Their numerical values will be a prediction of the fit. 

For the spectrum, we only assume the existence of a part of it. It turns out that to correctly reproduce the experimental data, we will need to consider at least four resonance zero constraints, one for the $\rho(770)$ in the $P$ wave, the two $f_0$'s in the $S_0$,\footnote{We call the resonances $f_0, f_0', f_0'',\dots$ to differentiate them. We will use similar notation in other channels. In the $S_0$ channel, we do not include the $\sigma$ in this list.} and one $f_2$ in the $S_2$ wave. We will then discover the spectrum by searching for additional dynamically generated zeros in the partial waves.

\subsection{The target functional}

\begin{figure}[t]
    \centering
    \includegraphics[width=\linewidth]{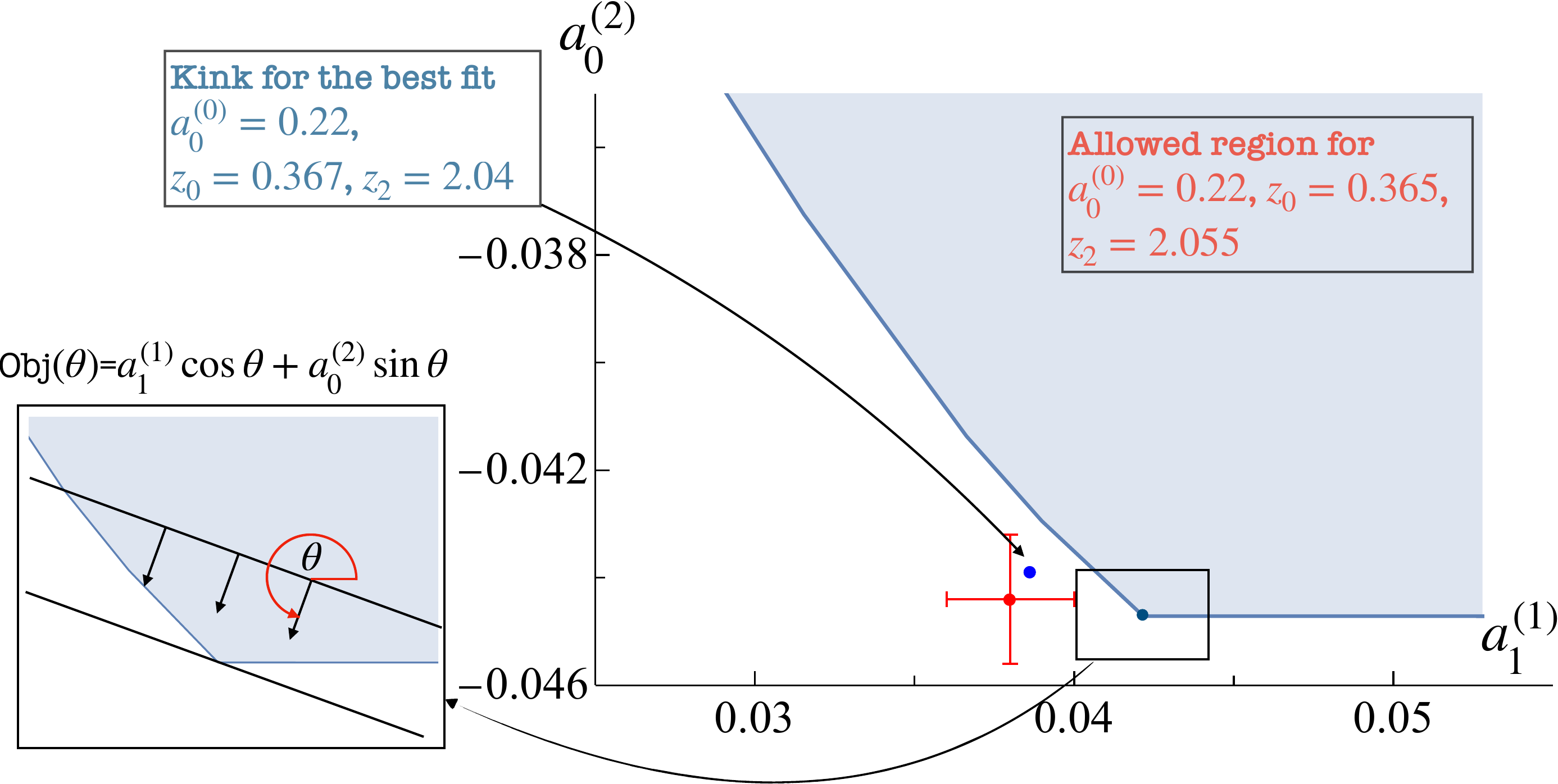}
    \caption{The kink from \cite{Andrea} in the scattering length space $\{a_0^{(0)}, a_1^{(1)}\}$. The allowed region is in blue. The point with error bars is the experimental determination, see also Table \ref{tab:thresholds_predictions}. The blue dot is the position of the kink as a result of our Bootstrap fit algorithm. On the left, we represent the target functional for $\theta=25\pi/18$.}
    \label{fig:kink}
\end{figure}

The final step in constructing our fit model is selecting the Bootstrap target functional. While there are, in principle, infinitely many possible choices, not all are equivalent. For our purposes, we found it convenient to adopt the approach developed in \cite{Andrea}.
To illustrate why, consider the most general ansatz
\eqref{multi-ansatz}. Assigning specific values to the $S_0$ scattering length\footnote{Scattering lengths are defined from the threshold expansion of the partial amplitudes $\Re t^{(I)}_\ell(s)=2 k^{2\ell}(a_\ell^{(I)}+k^2 b_\ell^{(I)}+\dots)$, where $k^2= s/4- 1$ is the center of mass momentum.} $a_0^{(0)}$
and the chiral zeros $z_0$ and $z_2$, we find that the allowed region for the $\{a_1^{(1)},a_0^{(2)}\}$ scattering lengths exhibits a kink, as shown in figure  \ref{fig:kink}. Changing the values of the $\{a_0^{(0)},z_0, z_2\}$ parameters, it is possible to move this kink closer to the experimentally measured scattering lengths (the red point in figure \ref{fig:kink}). When this occurs, the extremal amplitude at the kink closely resembles the pion amplitude. 

In this case, the scattering lengths and subleading threshold coefficients align well with experimental data, and the corresponding amplitude contains both the
$\sigma$, and the $\rho$ resonances. However, although the $\sigma$ position agrees with the data, the $\rho$ width is larger than observed experimentally.\footnote{We tried to fine-tune the triplet $\{a_0^{(0)},z_0, z_2\}$ to match the experimental values of $\sigma$ and $\rho$, but without success. We conjecture that by fine-tuning additional low energy constants it would be possible to generate the $\rho$ in the correct position dynamically.} This suggests that by adding the spectral assumption, it may be possible to develop a robust fit ansatz for the pion amplitude. 

 The Bootstrap problem formulated to construct the fit model for the pion amplitude can be summarized as follows.
 \begin{mdframed}[frametitle={Fit Ansatz},frametitlealignment=\centering,backgroundcolor=blue!6, leftmargin=0cm, rightmargin=0cm, topline=false,bottomline=false, leftline=false, rightline=false] 
\vspace{-0.4cm}
\begin{align} 
&\text{Given}
&& \Theta = \{\theta, a_0^{(0)},  z_0, z_2,  m_\rho^2,  m_{f_0}^2,  m_{f'_0}^2, m_{f_2}^2 \} \nn\\[5pt]
&\underset{\text{in } {\mathcal{A}^\text{ansatz}(s,t,u)}}{\text{Maximize}} &&  \text{Obj}(\theta)\nonumber\\
& \text{constr. by}  && t_0^{(0)}(4){=} 2 a_0^{(0)}, t_0^{(0)}( z_0){=}0, t_0^{(2)}( z_2){=}0 \nn\\ 
&& &S_1^{(1)}( m_\rho^2){=}0,S_0^{(0)}( m_{f_0}^2){=}0\nn\\
&& &S_0^{(0)}( m_{f_0^\prime}^2){=}0,S_2^{(0)}( m_{f_2}^2){=}0\nn \\
& s\geq 4&& \mathcal{U}_\ell^{(I)}\succeq 0 \,\, \text{for} \,\, \ell \in \mathbb{N}, \, I=0,1,2
 \label{eq:bootstrap_problem}
\end{align}
\end{mdframed}
The objective we maximize is given by the linear combination
\beq
\text{Obj}(\theta)=a_1^{(1)}\cos\theta+a_0^{(2)}\sin\theta,
\label{kink_objective}
\eeq
which depends on the angle $\theta$. This objective is a \emph{normal} functional and is effective at selecting ``kinks'' \cite{Cordova:2019lot}.  
To understand this, note that this objective maximizes the amplitude in the $\{a_0^{(0)}, a_1^{(1)}\}$ space at a point where the tangent vector is orthogonal to $(a_1^{(1)}\cos\theta,a_0^{(2)}\sin\theta)$. At a kink, where multiple tangents exist, many normal functionals naturally converge to the kink point.
For $\pi<\theta<3\pi/2$, this functional span the boundary of the allowed blue region in figure \ref{fig:kink}. 
In the left inset, we show a typical choice of angle $\theta$ that will select the kink in the $\{a_0^{(0)}, a_1^{(1)}\}$ space.\footnote{We could consider a different objective $\text{Obj}^\prime(\theta,\phi)=a_0^{(0)}\cos\phi \sin\theta +a_0^{(2)}\sin\phi\sin\theta+a_1^{(1)}\cos\theta$, and replace the $a_0^{(0)}$ parameter in \eqref{eq:bootstrap_problem} with a new angle $\phi$. This would lead to an equivalent formulation of the Bootstrap ansatz \eqref{eq:bootstrap_problem}. From this view, it is evident we are moving along the two-dimensional boundary of the three-dimensional parameter space of the scattering lengths. This can be generalized by adding an arbitrary number of low energy observables, which amounts to scanning over a larger space of amplitudes.} 

The quantities collectively denoted by $\Theta$ are the parameters of the fit and the input for the optimization problem \eqref{eq:bootstrap_problem}. In this problem, four parameters $\theta, a_0^{(0)}, z_0, z_2$ are real, while the mass square are complex. Their real and imaginary parts correspond respectively to the physical mass and width of the resonance. In total, the size of the parameter set in $|\Theta|=12$.
We solve this problem using the standard SDPB solver \cite{Simmons-Duffin:2015qma,Landry:2019qug}. The corresponding extremal amplitude depends on the choice of $\Theta$. Next, we will explain how to optimize on $\Theta$. 

\section{Gradient free optimization and particle swarm}
\label{sec:PSO}

For any choice of $\Theta$, the optimal solution of the Bootstrap problem \eqref{eq:bootstrap_problem} yields a model for the scattering amplitude $\mathcal{A}^\text{ansatz}_\Theta(s|t,u)$. To choose $\Theta$, we construct the $\chi^2(\Theta)$ using the Bootstrap amplitude and the experimental data, and we minimize it.
The dependence of the model $\mathcal{A}^\text{ansatz}_\Theta(s|t,u)$
is non-linear in $\Theta$, hence we expect the $\chi^2(\Theta)$ to be a non-convex function.

In this Letter, 
we explore an algorithm especially suited to this class of problems, the Particle Swarm Optimization algorithm (PSO) \cite{Shi1998AMP}.\footnote{We are grateful to Balt van Rees for pointing out this method.} 
The PSO is a standard algorithm designed for solving non-convex problems, with a wide range of applications---see \cite{PSOclerc} for an introduction. 
The PSO is also gradient-free and does not require the computation of derivatives of the Bootstrap solution in $\Theta$.\footnote{As explained in \cite{Reehorst:2021ykw}, it is possible to efficiently compute the gradient of a Bootstrap problem solved with the interior point method as the one implemented in SDPB.}

We start with $n_p$ particles at random positions $\Theta^{(i)}_0$ to which we assign random velocities $v^{(i)}_0$. 

At step $n$, we update their positions following the rule
\beqa
v^{(i)}_{n+1}&=&\omega v^{(i)}_n + c_1 r_1 (\Theta^{(i)}_n-X^{(i)}_n)+c_2 r_2 (\Theta^{(i)}_n - Y_n), \nn \\
\Theta^{(i)}_{n+1}&=&\Theta^{(i)}_n+v^{(i)}_{n+1}.
\label{eq:PSO_update}
\eeqa
The velocity of the $i$-th particle at step $n+1$ is thus a linear combination of its previous velocity, the distance to its position with the lowest $\chi^2$, denoted by $X_n^{(i)}$ and the distance with the overall best position among all particles $Y_n$. At each step, we first compute the $\chi^2(\Theta_n^{(i)})$ for each particle, then evaluate the various parameters to perform the step.

Three parameters control the algorithm performance: the inertia $\omega$, the cognitive coefficient $c_1$, and the cooperation coefficient $c_2$.\footnote{The $r_{1,2}$ are random numbers uniformly distributed between $[0,1]$, and are drawn at each step.} The convergence property of the algorithm depends on the choice of these three coefficients.\footnote{In other versions of the algorithm, those parameters can be promoted to be dynamical.} The details of our implementation of the PSO algorithm, which contains an additional adaptive velocity prescription as in \cite{Xu2013AnAP}, can be found in Appendix \ref{sec:swarm}. 

\section{Bootstrap Fit results}
\label{sec:best_fit}

\begin{figure*}[t]
    \centering
    \includegraphics[width=\linewidth]{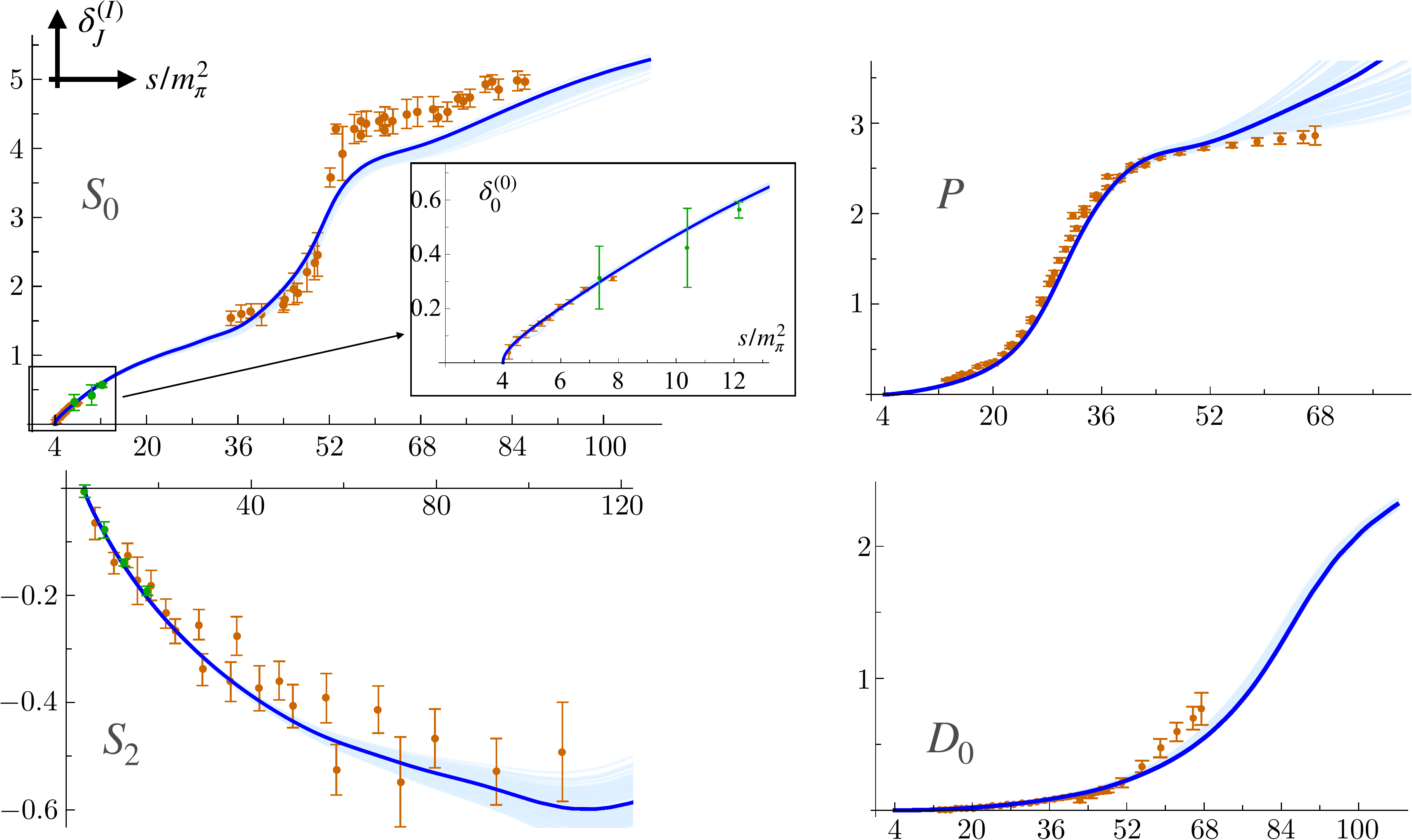}
    \caption{The four channels used to fit the pion amplitude. The points respectively in orange and green are experimental and lattice data. The thick blue curve is the best fit, the light blue cloud is given by all the curves with sub-optimal $\chi^2$. (Notation: $S_0$ stands for the $(I,\ell)=(0,0)$ channel, $P$ for the $(1,1)$, $S_2$ for the $(2,0)$, and $D_0$ for the $(0,2)$.}
    \label{fig:best_fit}
\end{figure*}

We construct the $\chi^2(\Theta)$
\beq\label{eq:chi2}
\chi^2(\Theta)=\sum_{i \in \text{data set}} \left(\frac{\delta_i^\text{exp}-\delta_\Theta^\text{Ansatz}(s_i)}{\Delta\delta_i^\text{exp}}\right)^2
\eeq
using experimental and lattice phase shifts $\delta_i^\text{exp}$, where the index $i$ stands collectively for the quantum numbers $(I,\ell)=\{(0,0),(1,1),(2,0),(0,2)\}$, and the energy $s_i$ of the measurement. Our ansatz for the phase shift is obtained by projecting $\mathcal{A}^\text{ansatz}_\Theta$ in partial waves and using the definition $S_\Theta=|S_\Theta|\, \text{exp}(2i\delta_\Theta)$.

The input data used in \eqref{eq:chi2} are from \cite{Protopopescu:1973sh,Estabrooks:1974vu,Grayer:1974cr,Losty:1973et,Hoogland:1977kt}, except for the $S_0$ wave. The data in the $S_0$ channel above $s\approx 20$ extracted from the old experiments are often incompatible and suffer from unknown systematic errors. Following \cite{Pelaez:2004vs}, we used only a selection of data points.\footnote{We are grateful to Jose Ramon Pelaez for highlighting this point.}
Close to the threshold, the experimental situation was cleared by the CERN experiment NA48/2 \cite{Batley:2010zza}. We also include lattice data from the RBC and UKQCD collaborations in the $S_0$ and $S_2$ channels extrapolated at the physical pion mass \cite{RBC:2021acc}.\footnote{For the lattice data points, the errors on the phase shift and the energy are strongly correlated. We do not include this effect in this work. In this case, we use the isospin symmetric value of the pion mass $m_\pi=135$ MeV to express energies in dimensionless units.}

The result of the Bootstrap fit is shown in figure \ref{fig:best_fit}. 
The orange data points are experimental, the green are taken from the lattice. The dark blue curve is the best fit with minimum $\chi^2(\Theta)\approx 40$. 
We do not assign a statistical meaning to the value of the $\chi^2$, but take it as a likelihood measure. The largest contribution to the $\chi^2$ comes from the $P$ wave, with $\chi^2\approx 30$, where the experimental error is almost negligible. We plan to investigate this channel with more care in the future, replacing the phase shift data with the determination from the pion form factors \cite{deTroconiz:2004yzs,Colangelo:2018mtw}, and with lattice extrapolations \cite{Boyle:2024grr}. 
The second largest contribution of order ten is due to the $S_0$ wave. Here we observe a systematic error in reproducing the phase above the $K\bar K$ threshold $s\approx 50$ (the $\chi^2$ below this energy is of order one). We believe this is due to the lack of an explicit threshold in our ansatz at that scale, as corroborated by the slower convergence of the Bootstrap problem, see also Appendix \ref{sec:numerical_details}.
Both $S_2$ and $D_0$ waves have $\chi^2$ of order one.

We run the PSO algorithm for $N_\text{iter}=60$ steps, using $n_p=10$ particles. We explore 600 points in the $\Theta$ parameter space and at each point, we construct the full analytic amplitude solving the Bootstrap problem \eqref{eq:bootstrap_problem}. To estimate errors we consider a subset of amplitudes, and compute the weighted average using the value of the $\chi^2$. We observe that 50\% of our set have $\chi^2<100$, and 15\% have $\chi^2<50$. We compare the errors estimated using these two cutoffs and find that those obtained with the looser one are a factor of two larger. The light-blue curves in figure \ref{fig:best_fit} are drawn from amplitudes with $\chi^2(\Theta)<50$. In the remainder of this paper, we will follow the same color scheme.

In the inset in figure \ref{fig:best_fit}, we zoom in on the NA48/2 and lattice data close to the threshold. The point at the kaon mass $s\approx 12$ is from lattice \cite{RBC:2021acc}. This point is crucial to help stabilize the amplitude between the threshold and the cluster of points around $s\approx 40$.

\begin{table}[h]
\centering
\begin{tabular}{||c | c  | c||} 
 \hline
  $\Theta$ &  Bootstrap Fit  & Literature  \\ [0.5ex] 
 \hline 
 \hline
\,  $a_0^{(0)}$  \, &  \quad $0.217 \pm 0.002$ \quad & $0.2196\pm 0.0034$ \cite{Batley:2010zza}  \\
 \hline
\,   $z_0$ \, & \quad $0.368\pm 0.008$ \quad &    \\
 \hline
\,   $z_2$ \, & \quad $2.040\pm 0.004$ \quad &   \\
 \hline
 \,   $m_\rho$ \, &($5.546 \pm 0.005$)&  $(5.555\pm 0.015)$ \\ & +$i(0.538{\pm} 0.002)$ &
 +$i(0.528\pm 0.013)$
 \\
 \hline
  \,   $m_{f_0}$ \, &$(7.18{\pm} 0.04)$+$i(0.26{\pm} 0.02)$& $(7.25{\pm}0.11)$+$i(0.21{\pm} 0.07)$ \\
  \hline
  \,   $m_{f_0^\prime}$ \, & $(9.8\pm 0.2)$+$i(1.7\pm 0.1)$ & $(9.8\pm 0.7)$+$i(1.3\pm0.9)$  \\
  \hline
   \,  $m_{f_2}$ \, & $(9.26{\pm} 0.03)$+$i(0.69{\pm} 0.04)$ & $(9.26{\pm}0.08)$+$i(0.73{\pm}0.08)$ \\
  \hline
 \,   $\theta/\pi$ \, & \quad $1.328 \pm 0.026$ \quad & \\
 \hline
\end{tabular}
\caption{Our estimate of the fit parameter compared and corresponding values quoted in the literature. Experimental values of the resonances are given by the PDG average \cite{ParticleDataGroup:2024cfk}.}
\label{tab:fit_parameters}
\end{table}

In Table \ref{tab:fit_parameters}, we list our estimate for the fit parameters and the corresponding errors. 
The error in this table is conservative and estimated using all the points with $\chi^2(\Theta)<100$ corresponding to half of the whole dataset of amplitudes produced. 

Beyond the statistical error, there are three more sources of systematics. The first concerns the convergence of the Bootstrap model $\mathcal{A}^\text{ansatz}_\Theta(s|t,u)$ in equation \eqref{multi-ansatz} that depends on $N_\text{vars}$ the number of free variables $\alpha_{n,m}^{(\sigma)}$, and $\beta_{n,m}^{(\sigma)}$. We used two ansatzes respectively with $N_\text{vars}=397$ and $N_\text{vars}=547$. All the results we quote in this paper are taken from the latter. The difference between the fit parameters estimated from the two ansatzes is comparable with the statistical error unless stated otherwise. The second source comes from the PSO search algorithm, which does not guarantee finding the global minimum. We checked the stability of our fit by performing different searches varying the search parameters that produced almost the same result. The last source of systematics comes from the choice of the Bootstrap functional. In this case, we have tested an alternative Bootstrap formulation obtaining the same fit.

\section{Predictions}
\label{sec:predictions}

The big advantage of our strategy is that the result of the fit is a full analytic amplitude. Next, we study its properties away from the region directly constrained by the fit.

\subsection{Amplitude below threshold}

\begin{figure}[t]
    \centering
    \includegraphics[width= \linewidth]{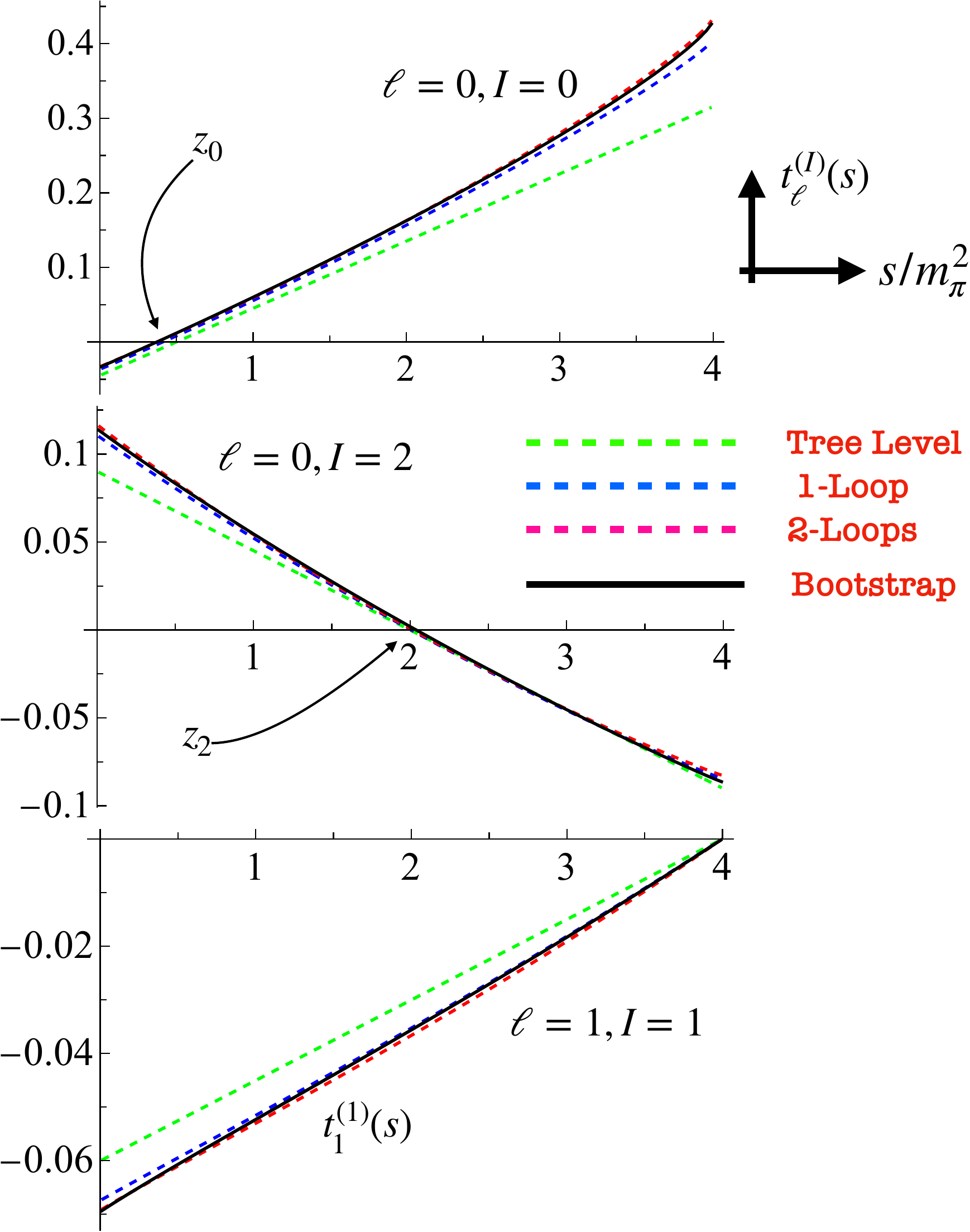}
    \caption{Partial wave amplitudes $t_\ell^{(I)}$ in the real strip $0<s<4$. The black line is the Bootstrap prediction, dashed lines are different approximations from $\chi$PT \cite{Bijnens:1995yn}.}
    \label{fig:ChiPT}
\end{figure}

The $\chi$PT expansion is a reliable approximation of the pion amplitude at low energy. In figure \ref{fig:ChiPT}, we plot the partial waves $t_0^{(0)}$, $t_0^{(2)}$, and $t_1^{(1)}$ of our best fit (the solid black line) in the sub-threshold region $0<s<4$, and we compare it with perturbation theory \cite{Bijnens:1995yn}. The tree level, denoted with the dashed green line is given by 
\beq
t_0^{(0)}=\frac{2s-1}{32 \pi f_\pi^2},\quad  t_1^{(1)}=\frac{s-4}{96\pi f_\pi^2},\quad t_0^{(2)}=\frac{2-s}{16\pi f_\pi^2}\,,
\eeq
where $f_\pi$ is the pion decay constant.
The higher loop expressions are more involved and are also plotted in figure \ref{fig:ChiPT}.
 The two chiral zeros $z_0$, and $z_2$  are also visible in the figure. We emphasize that the only input in this region is the existence of the two chiral zeros, not their position. 

Next, we extract the threshold parameters from our amplitude and compare them with the values of  \cite{Colangelo:2001df} in Table \ref{tab:thresholds_predictions}. 
We find a very nice agreement for the leading and sub-leading threshold coefficients for all waves below $\ell=2$. The scattering lengths we extract are different for higher spins but of the same order. It would be interesting to investigate the reason for this discrepancy.

\begin{table}[h]
\centering
\begin{tabular}{||c | c  | c||} 
 \hline
  \quad &  Bootstrap Fit  & Literature  \\ [0.5ex] 
 \hline 
 \hline
  $a_0^{(2)}$  
  &  \, $(-0.432\pm 0.001)\times 10^{-1}$ \, & $(-0.444\pm 0.012)\times 10^{-1}$   \\
 \hline
   $a_1^{(1)}$  &  $(0.380\pm 0.002)\times 10^{-1}$  &  $(0.379\pm 0.05)\times 10^{-1}$  \\
 \hline
   $b_0^{(0)}$  & $ 0.265 \pm 0.030$  & $0.276\pm 0.006$  \\
 \hline
    $b_0^{(2)}$  & $(-0.797\pm 0.002)\times 10^{-1}$ &  $(-0.803\pm0.012)\times 10^{-1}$ \\
 \hline
     $b_1^{(1)}$  & $(0.61 \pm 0.02)\times 10^{-2}$ & $(0.57 \pm 0.01)\times 10^{-2}$\\
  \hline
     $a_2^{(0)}$  
     & $(0.53\pm 0.11)\times 10^{-2}$ & $(0.175\pm 0.003)\times 10^{-2}$ \\
  \hline
    $a_2^{(2)}$  & $(0.51 \pm 0.18)\times 10^{-3}$ & $(0.170\pm 0.013)\times 10^{-3}$\\
  \hline
    $a_3^{(1)}$ & $(1.5\pm 0.4)\times 10^{-4}$  & $(0.56\pm 0.02)\times 10^{-4}$\\
 \hline
\end{tabular}
\caption{Except for the values of $a_0^{(2)}$ taken from \cite{Batley:2010zza}, all other threshold parameters are taken from \cite{Colangelo:2001df}. }
\label{tab:thresholds_predictions}
\end{table}

\subsection{The $D_2$ partial wave}

\begin{figure}[t]
    \centering
    \includegraphics[width=0.9\linewidth]{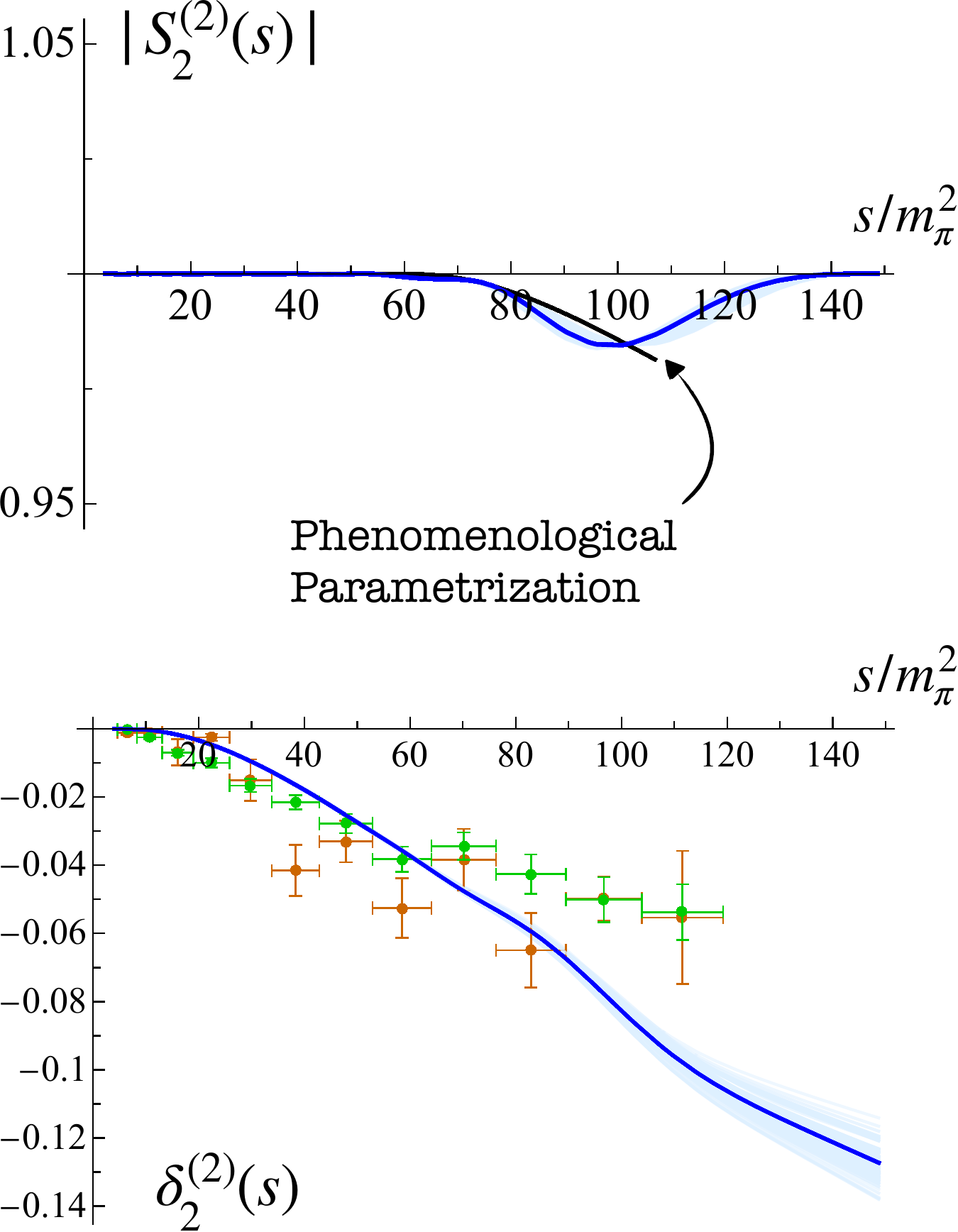}
    
\caption{
Top panel: Elasticity of the $D_2$ wave as a function of $s/m_\pi^2$. The black line represents the parametrization from \cite{Pelaez:2004vs}.  
Bottom panel: Phase shift of the $D_2$ wave compared with the available experimental data. The green and orange points correspond to the two possible phase determinations reported in \cite{Hoogland:1977kt}.}

    \label{fig:S22}
\end{figure}

In figure \ref{fig:S22}, we plot the elasticity $|S_2^{(2)}|$ and the  $\delta_2^{(2)}$ phase shift prediction. The color scheme follows the one defined in figure \ref{fig:best_fit}.
The data for the phase shift are taken from \cite{Hoogland:1977kt}, orange and green correspond to two possible determinations. 
We compare the elasticity profile with the phenomenological parametrization from \cite{Pelaez:2004vs}. It is interesting to notice a dip in the unitarity around $s\approx 90$. The dip is insensitive to Bootstrap systematics.
We think that the emergent particle production in this channel is a consequence of the inelasticity profile injected in the $D_0$ wave. This effect is compatible with our expectation that including mixed amplitudes with kaons and pions might lead to a correct prediction for the inelasticity.

\subsection{Spectrum}
\label{sec:spectrum}
Beyond the spectrum assumed and fitted using the experimental data, we also observe several zeros dynamically generated in our construction.

\begin{figure}[t]
    \centering
    \includegraphics[width=\linewidth]{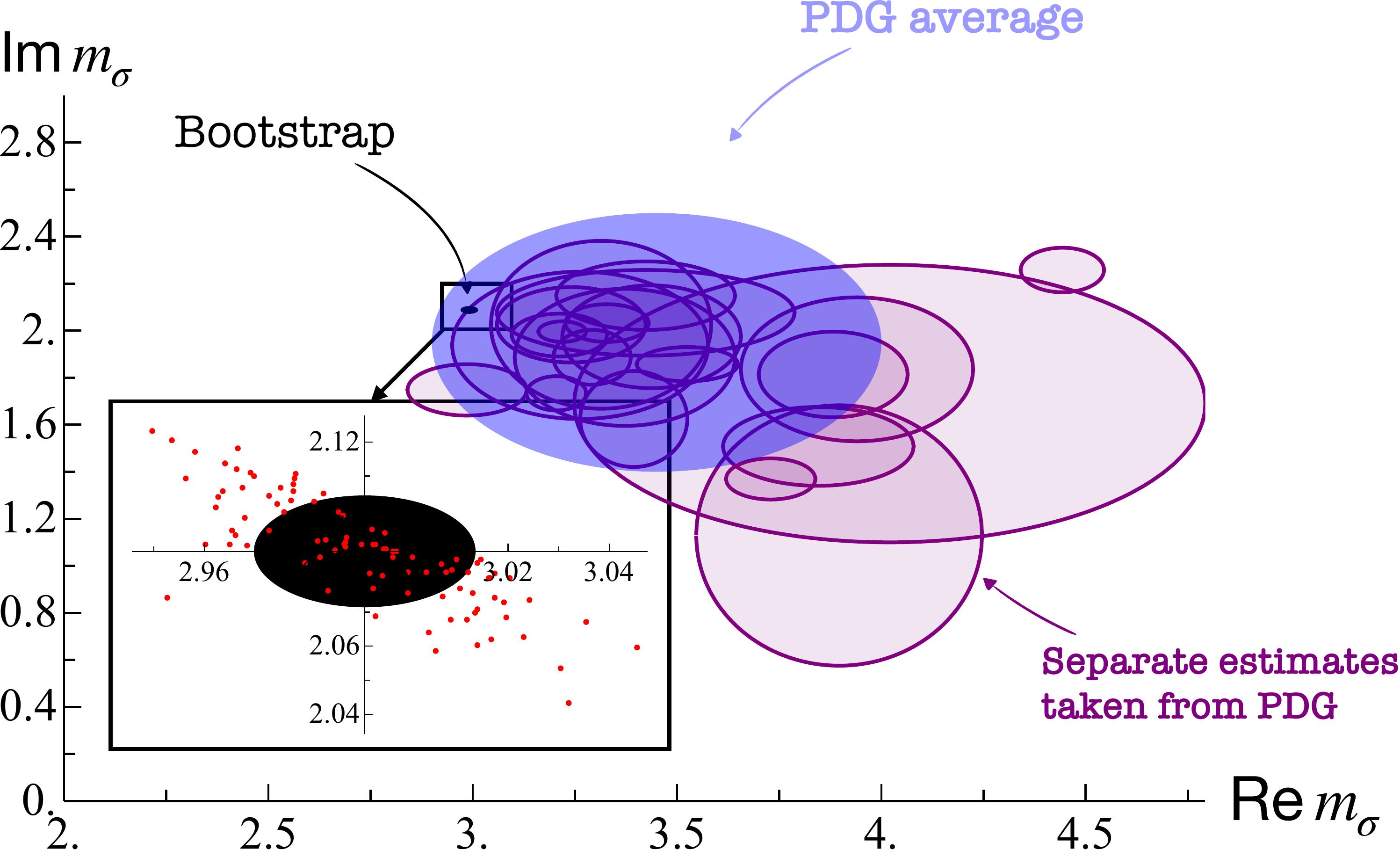}
    \caption{Determinations of real $\Re m_\sigma$ and imaginary part $\Im m_\sigma$ of the mass of the $\sigma$. The black ellipse highlighted in the inset is our Bootstrap determination.}
    \label{fig:Sigma}
\end{figure}

We begin by discussing the $f_0(500)$ resonance, commonly referred to as the $\sigma$. We summarize the phenomenological determinations of its position in figure \ref{fig:Sigma}. The pink ellipses denote all the estimates of its complex mass reported in the PDG since 2001 \cite{ParticleDataGroup:2024cfk}. The blue ellipse is the PDG average. Our estimate is represented by the black ellipse highlighted in the inset.

\begin{figure}[t]
    \centering
    \includegraphics[width=\linewidth]{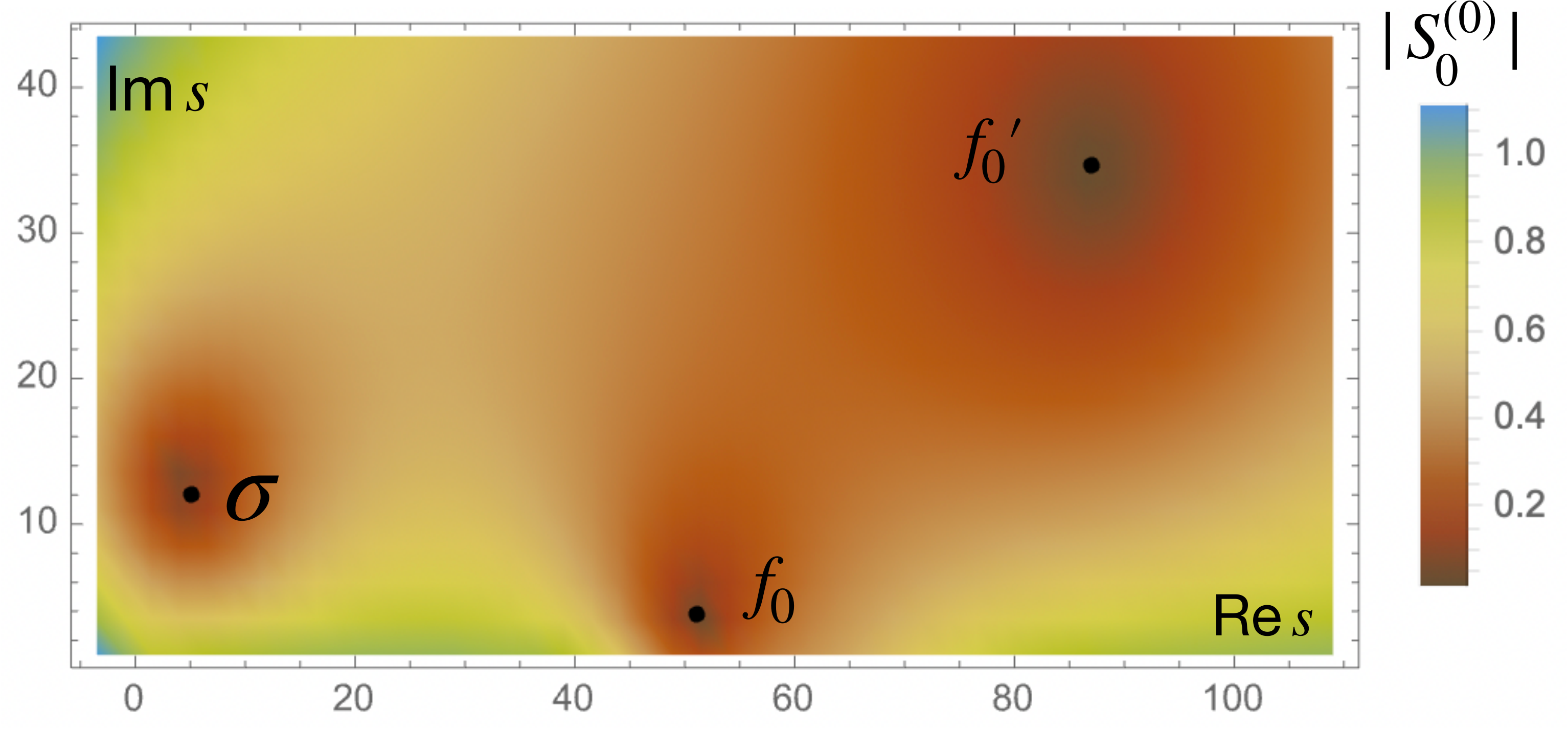}
    \caption{$|S_0^{(0)}|$ in the upper half complex $s$ plane plotted for the best-fit amplitude. We highlight with black dots the three scalar resonances found in this channel.}
    \label{fig:Sigma_Plane}
\end{figure}

To estimate the position of the $\sigma$ we look for a zero at low energy in the $S_0^{(0)}$ $S$-matrix using the Newton method. We repeat the operations for all amplitudes with $\chi^2(\Theta)<50$, and perform a weighted average to predict the mass and determine the error. The red dots in the inset are the individual determinations from this sample.
Our final estimate is reported in Table \ref{tab:dynamic_resonances}.
In figure \ref{fig:Sigma_Plane}, we also show the density plot for $|S_0^{(0)}|$ in the upper-half complex $s$ plane. The left-most zero is the $\sigma$. In the same plot, we observe two additional resonances that we identify as the $f_0(980)$, and $f_0(1370)$. 

Our $\sigma$ determination is shifted away from the center of the PDG average. This is correlated with the lattice point at $s=m_K^2 \approx 12$ which was not used in previous studies, and impacts the growth of the phase in the $S_0$ wave. 

\begin{table}[h]
\centering
\begin{tabular}{||c | c  | c||} 
 \hline
   &  Prediction  & PDG average  \\ [0.5ex] 
 \hline 
 \hline
 \,   $m_\sigma$ \, &$(2.99{\pm} 0.02)$+$i(2.09{\pm} 0.02)$&  $(4.3\pm 1.5)$+$i(3.3\pm 2.6)$ \\
 \hline
  \,   $m_{\rho^\prime}$ \, &$(9.5\pm0.6)$+$i(0.55\pm0.05)$& $(10.7\pm 0.3)$+$i(1.45\pm 0.05)$ \\
  \hline
  \,   $m_{\rho_3}$ \, & $(16.7\pm 0.6)$+$i(0.7\pm 0.2)$ & $(12.3\pm 0.1)$+$i(0.68\pm 0.05)$  \\
  \hline
   \,  $m_{\mathbb{T}}$ \, & $(14.6\pm 0.3)$+$i(2.4\pm 0.1)$ & ? \\
 \hline
\end{tabular}
\caption{Experimental values of the resonances are taken from \cite{ParticleDataGroup:2024cfk}.}
\label{tab:dynamic_resonances}
\end{table}

\begin{figure}[t]
    \centering
    \includegraphics[width=\linewidth]{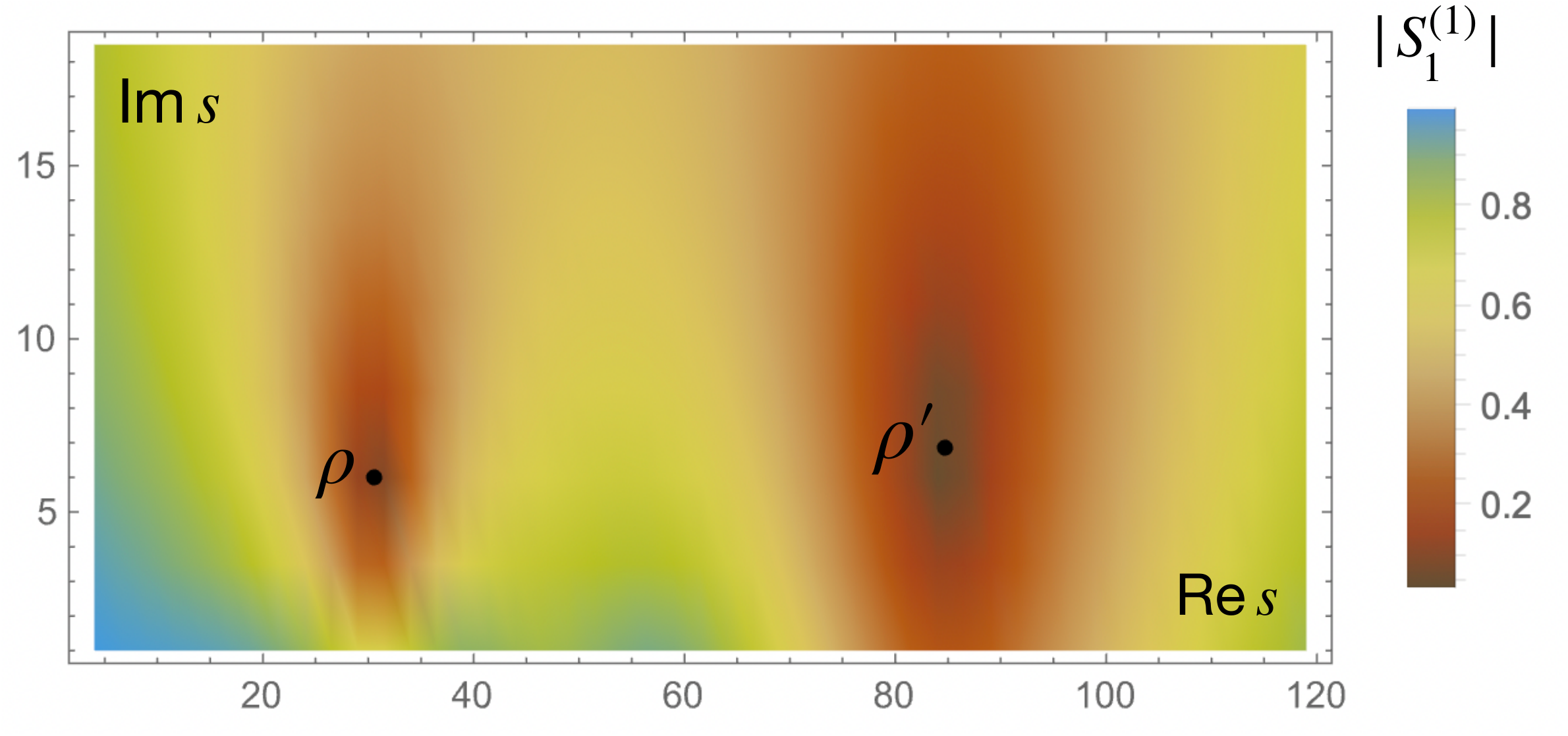}
    \caption{$|S_1^{(1)}|$ density profile in the upper half complex $s$ plane. We highlight with black dots the two $\rho$ resonances in our amplitude.}
    \label{fig:Rho_Plane}
\end{figure}

We also find an additional state in the $P$-wave. The density plot for the $|S_1^{(1)}|$ in the upper half $s$ plane is shown in figure~\ref{fig:Rho_Plane} for the best-fit amplitude. The two black dots are zeros of $S_1^{((1)}$. The left-most zero is identified as the $\rho(770)$ resonance.  Its position is fixed by the fitting procedure. The second black dot on the right is dynamically generated and has a mass 5\% lower than the experimental determination of the $\rho(1450)$. However, its width is half of the experimentally measured value. This resonance still suffers from systematics: we do not have the inelasticity profile at this energy, and $N_\text{vars}$ of the Bootstrap problem is not large enough to accurately describe this region. We suspect that dealing with this systematics might lead to the correct width in our procedure.

For higher spins, our amplitude dynamically generates a spin $\ell=3$ resonance, this is the first sign of Reggeization. This is typical in the amplitudes constructed using the S-matrix Bootstrap where the high energy behavior instead of being erratic is physical, although it converges slowly, see also section \ref{sec:high-energy}. The mass of the $\rho_3$ obtained is $20$\% above the experimental value, and its width is larger. This resonance, however, is still affected by the convergence of the Bootstrap ansatz, and improving the numerics might lead to a better agreement.\footnote{The other possibility is that to obtain the precise Regge trajectory we might need additional fine-tuning, which can be realized by enlarging the set of fit parameters or generalizing the Bootstrap problem we use to construct the amplitude.}

\subsection{The tetraquark}

\begin{figure}[t]
    \centering
    \includegraphics[width= \linewidth]{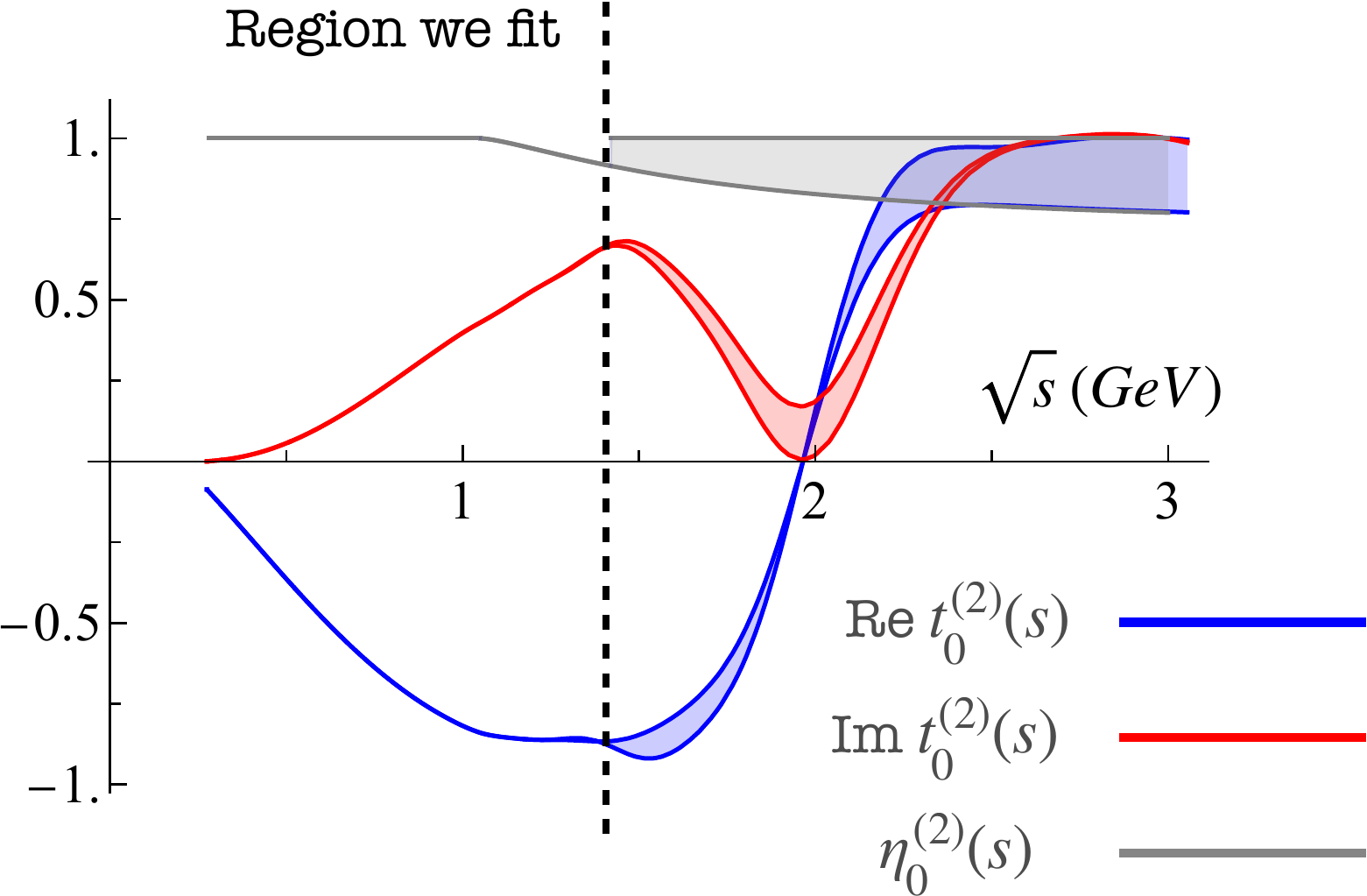}
    \caption{Real and imaginary parts of the amplitude in the 
 $S_0^{(2)}$ channel. The error bars indicate the systematic uncertainties in the line shape of this function arising from the unknown inelasticity above 1.4 GeV. This error is comparable to the statistical error of our fit reported in figure \ref{fig:Tetra_Phase}.
}
    \label{fig:LShape}
\end{figure}

As already indicated in the introduction, examining the $S_2$ wave reveals an unexpected result. Continuing the phase shift $\delta_0^{(2)}$ to high energy, as shown in figure \ref{fig:Tetra_Phase}, we observe a broad resonance with isospin $I=2$. 
This indicates the presence of a genuine Tetraquark state \cite{Jaffe:1976ig,tHooft:2008rus}. In the physical world, where electric charge is reintroduced, such particles would carry a charge of two.
Its mass and width in physical units are respectively 2 GeV and 600 MeV. It would be intriguing to examine the invariant mass distributions of $\pi^{\pm}\pi^{\pm}$ in 
$B$-meson decays that provide sufficient energy to reach the 2 GeV scale. Examples include the decay $B^+\to \pi^-\pi^+\pi^+$ \cite{LHCb:2019sus}, which can be studied at LHCb, as well as similar decays where the $\pi^-$  is replaced by either $D^-$ or $K^-$.

We determine the mass and width of the Tetraquark by locating the zero in the complex $s$ plane in the $S_2$ wave. This determination is free from Bootstrap-related systematics, as the amplitude in this channel is well-converged. However, we have not accounted for inelastic effects in this region, which could slightly alter the resonance's position. 

In figure \ref{fig:LShape}, we plot the real and imaginary parts of the partial wave amplitude $t_0^{(2)}$ as a function of $\sqrt{s}$ measured in GeV. From an amplitude perspective, this represents the experimental signal expected in the presence of a Tetraquark.\footnote{We thank Alessandro Pilloni for providing insights into the input required for the experimental data analysis.} The black dashed line separates the region where we fit the data and incorporates the inelasticity profile from the high-energy region, which is unconstrained. The gray curve represents the elasticity profile $\eta_0^{(2)}(s)$ taken from \cite{Garcia-Martin:2011iqs}. Above $\sqrt{s}=1.4$ GeV, we set $\eta_0^{(2)}(s)=1$ in our procedure. For illustrative purposes, we show the amplitude profile using the 
 $S_2$ wave parametrization $S_0^{(2)}(s)=|S_0^{(2)}(s)| \exp(2i\delta_0^{(2)}(s))$, where $\delta_0^{(2)}$ is the phase obtained from our best fit, and $|S_0^{(2)}(s)|$ is a generic function bounded by $\eta_0^{(2)}(s) \leq |S_0^{(2)}(s)| \leq 1$, for $s>1.4$ GeV.
 The lower bound corresponds to the extrapolation of \cite{Garcia-Martin:2011iqs} to higher energies, as shown by the gray area in figure \ref{fig:LShape}. The blue and red bands depict how uncertainty in the elasticity function influences the shape of the amplitude.

It would be interesting to understand the mechanism behind the emergence of the Tetraquark, using, for instance, the Roy equations \cite{Roy:1971tc}. Notably, there is a two-dimensional example that qualitatively mirrors this scenario: the two-dimensional theory of the QCD flux tube. In this context, the QCD $\rho$ is analogous to the axion, while the $\sigma$ corresponds to the dilaton. The analogy is striking, as this model features a sharp axion and a broad dilaton \cite{Gaikwad:2023hof}. Furthermore, in \cite{FluxTube}, it was observed that crossing symmetry necessitates a broad resonance in the symmetric channel. A similar mechanism might underlie the presence of the Tetraquark identified in our amplitude.

\subsection{High Energy}
\label{sec:high-energy}

\begin{figure*}[t]
    \centering   \includegraphics[width=\linewidth]{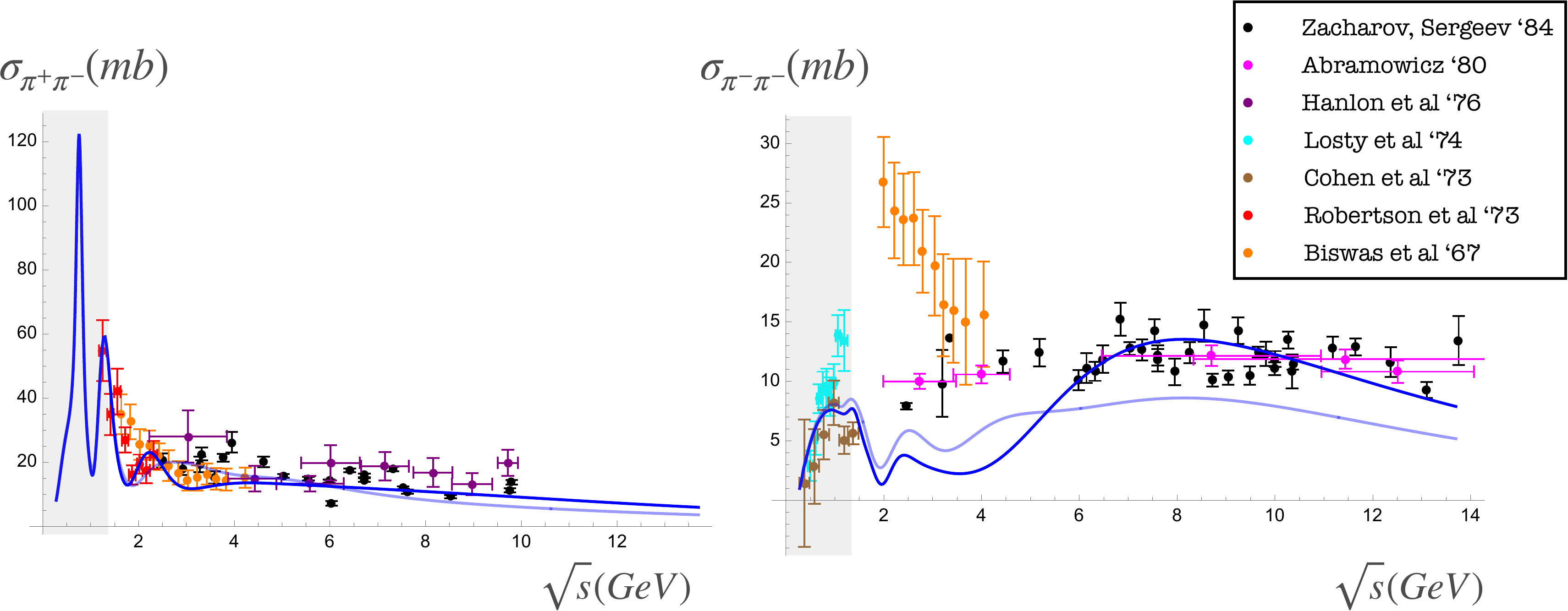}
    \caption{Total cross sections for $\pi^+\pi^-\to \pi^+\pi^-$ and $\pi^-\pi^-\to\pi^-\pi^-$. The gray shaded area is the region where we fit the data. The blue curve is the best fit for $N_\text{vars}=547$, in light blue the best fit for $N_\text{vars}=397$. Above 2 GeV there is still room to improve numerics, but there is already a very good agreement between Bootstrap and the experimental data \cite{Biswas:1967mpl,Robertson:1973tk,Cohen:1973yx, Losty:1973et, Hanlon:1976ct, Abramowicz:1979ca, Zakharov:1984nq}.}
    \label{fig:high_energy}
\end{figure*}

Finally, we examine the Regge behavior of our amplitude and compare it with experimental data.

One of the most intriguing recent insights into S-matrix Bootstrap amplitudes is their emergent high-energy behavior. Although the Bootstrap ansatz in \eqref{multi-ansatz} approaches a constant at infinity, there exists an intermediate regime where the numerical amplitudes align with Regge theory.
Specifically, at large 
$s$ and fixed $t$, the amplitudes exhibit the expected behavior $\mathcal{A}\sim s^{\alpha(t)}$
  \cite{Bhat:2023puy, Gumus:2023xbs, Reggeworkinprogress, froissartbound}.

In figure \ref{fig:high_energy}, we compare the experimental total cross sections for $\pi^+\pi^-$ and $\pi^- \pi^-$ with those extracted from our best-fit amplitude, shown in blue\footnote{In isospin components $\sigma_{\pi^+\pi^-}\propto \tfrac{1}{6}(2\mathcal{T}^{(0)}+3\mathcal{T}^{(1)}+    \mathcal{T}^{(2)})$, and $\sigma_{\pi^-\pi^-}\propto \mathcal{T}^{(2)}$.}
\beqa
\sigma_{\pi^+\pi^-}(s)&=& \frac{\Im\left[\mathcal{A}(s|t,u)+\mathcal{A}(t|s,u)\right]}{\sqrt{s(s{-}4)}}\biggr|_{t=0},\nn \\
\sigma^{\pi^-\pi^-}(s)&=&\frac{\Im\left[\mathcal{A}(t|s,u)+\mathcal{A}(u|s,t)\right]}{\sqrt{s(s-4)}}\biggr|_{t=0},
\eeqa
The solid blue curve represents our fit with the highest 
$N_\text{vars}=
547$, while the light blue curve corresponds to the best fit with 
$N_\text{vars}=397$.

In $\sigma_{\pi^+\pi^-}$
we observe distinct peak structures
corresponding to various resonances in the spectrum, with the first and most prominent being the $\rho$ peak. In contrast, $\sigma_{\pi^-\pi^-}$ is not expected to exhibit resonance-related peaks, except for the Tetraquark. At asymptotically high energies, we expect a constant or slightly growing behavior consistent with the pomeron exchange (see for instance \cite{Caprini:2011ky}). As $N_\text{vars}$  increases, we expect the Bootstrap amplitude will better approximate this behavior, which is not explicitly enforced in our ansatz \eqref{multi-ansatz}.

\section{Discussion and outlook}
\label{sec:discussion}

This Letter introduces a novel constrained approach for fitting experimental data. The key advantage of our method lies in its control over theoretical approximations. The amplitude model employed is a genuinely analytic, crossing-symmetric function that satisfies non-perturbative unitarity. The algorithm combines semi-definite programming and non-convex optimization techniques: for the Bootstrap component, we use SDPB \cite{Simmons-Duffin:2015qma,Landry:2019qug}, while for the particle swarm optimization (PSO), we implemented a straightforward Mathematica notebook.

The constructed $\pi\pi\to\pi\pi$ amplitude exhibits several remarkable features. It reproduces the experimental data, accurately predicts the $G$-parity plus spectrum below 1.4 GeV, aligns with $\chi$PT at low energies, and qualitatively matches the expected behavior in the high-energy regime.
Additionally, the model predicts a resonance heavier than the 
$\rho$ in the 
$P$ wave, with a mass comparable to the real-world $\rho(1450)$, as well as a spin-three resonance approximately 20\% heavier than the experimental value. 
The $D_2$ phase shift and inelasticity also show good agreement with experimental data and phenomenological models. 

\begin{figure}[t]
    \centering
    \includegraphics[width=\linewidth]{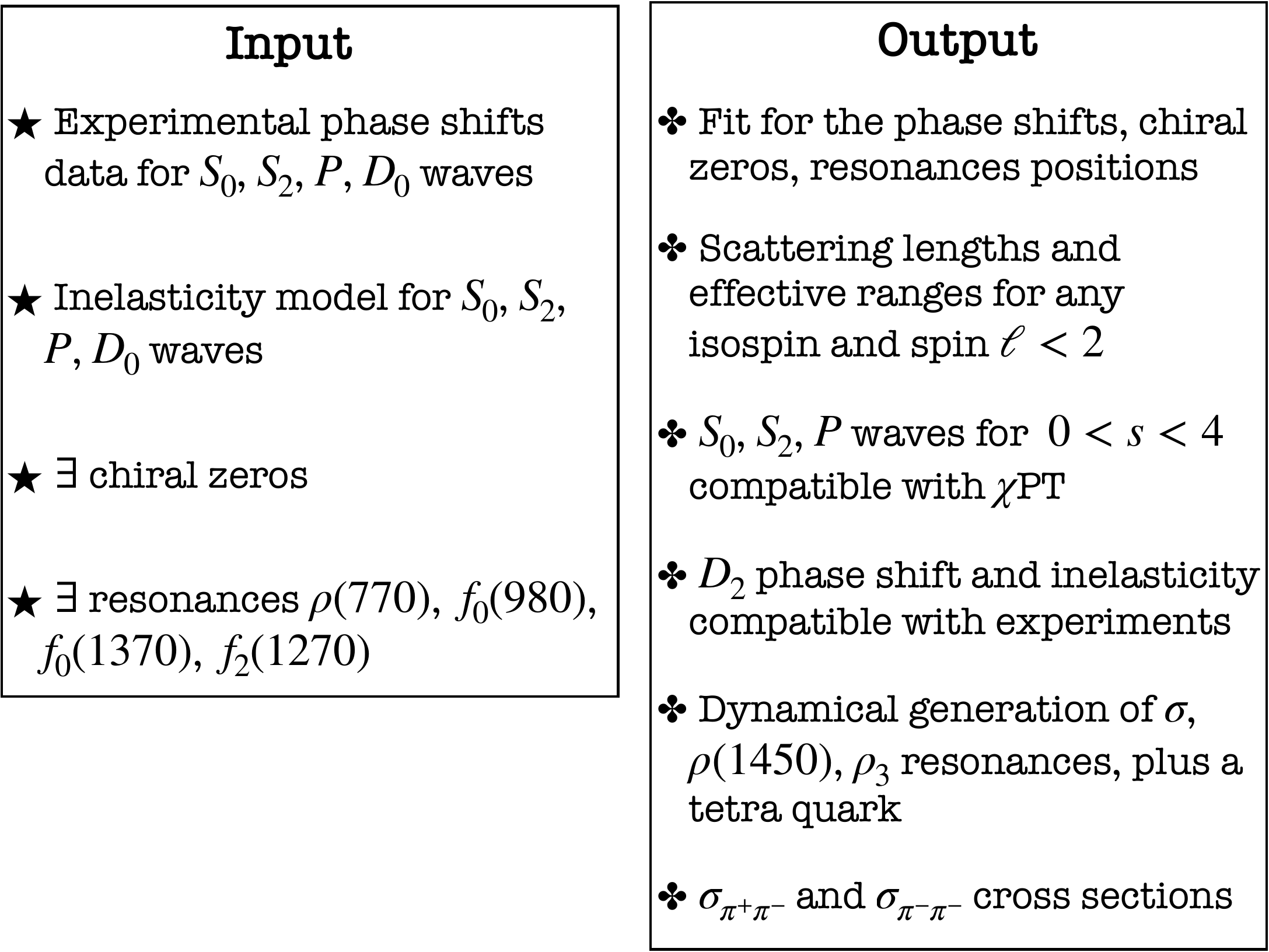}
    \caption{Summary of the bootstrap fit procedure.}
    \label{fig:summary}
\end{figure}

Interestingly, our analysis predicts a tetraquark state with a mass of approximately 2 GeV in the $I=2$ channel. This is a nonperturbative prediction, and to our knowledge, no other model makes a similar claim. Given that the constructed amplitude accurately describes and predicts several key features of $\pi\pi$ scattering, we are confident in the existence of this particle. Investigating its presence in the decays of 
$B$ mesons would be particularly intriguing.

Several enhancements could further improve the pion amplitude model we have constructed.

First, incorporating the correlation matrix for the lattice data in the $S_0$ wave from \cite{RBC:2021acc} would be valuable for assessing its impact on the determination of the $f_0(500)$ mass parameters. In particular, the lattice data are the source of the small tension with the previous $\sigma$ determination from Roy equations \cite{Caprini:2005zr}.
For the $P$ wave, it would be worthwhile to include the $\delta_1^{(1)}$ phase extracted from form factors \cite{deTroconiz:2004yzs,Colangelo:2018mtw} as well as results from lattice QCD extrapolations at the physical pion mass \cite{Boyle:2024grr}. 

The functional \eqref{kink_objective} used to construct our fit model might appear ad hoc and challenging to generalize to other physical systems.
To address this, we plan to explore alternative optimization approaches in future work \cite{fit2_work_in_progress}. The availability of such alternative constructions could also introduce a source of systematic uncertainty that should be incorporated into our estimates---something not currently accounted for.\footnote{We have already tested an alternative fitting approach, which yields nearly identical results. It would also be interesting to understand the relation between this objective and the regions selected in \cite{Bose:2020shm,Bose:2020cod}, and \cite{He:2023lyy, He:2024nwd} using alternative principles.}

Another source of systematic uncertainty in our fit procedure is the input for the shape of the inelasticity. To make the fit more self-consistent, we could expand the set of amplitudes to include mixed processes involving pions and kaons. This would provide access to additional resonances and allow us to incorporate more experimental data. The inelasticity profile could then be treated as an output of the fit.

Moreover, we believe that replacing the spectrum assumption and dynamically generating all resonances would be ideal.
This could be achieved by optimizing larger linear combinations of low-energy constants and generalizing the functional in equation \eqref{kink_objective} as has been done in several S-matrix Bootstrap studies \cite{EliasMiro:2019kyf, Guerrieri:2020bto, 
Bose:2020cod,
Guerrieri:2021ivu, Guerrieri:2022sod,Haring:2022sdp, Acanfora:2023axz}. 

It would also be interesting to analytically continue the best-fit amplitude in spin, as done in \cite{Acanfora:2023axz}, and explore the Regge trajectories for complex spins. 
This would necessitate using a larger ansatz to ensure sufficient numerical precision. We could then investigate in detail the resonance spectrum, and understand whether the three $0^+(0^{++})$ scalar resonances below 1.4 GeV are subtraction poles or part of Regge trajectories.

Our amplitude could also be used to measure form factors, as explored in \cite{Karateev:2019ymz, Correia:2022dyp, Cordova:2023wjp}. By fixing our parametrization, we could extract the form factors as an output and verify their consistency with high-energy sum rules recently explored in \cite{He:2023lyy, He:2024nwd} for $\pi\pi$ scattering.\footnote{Form factors computed using other methods, such as in \cite{Chen:2021pgx}, where Hamiltonian truncation was employed, could also be used to constrain the amplitude.}

Ultimately, how can we extend our method and determine rigorous bounds on the position of the resonances in 
figure \ref{fig:spectrum} determined from the Bootstrap Fit procedure? 
In this direction, it would be interesting to generalize the strategy proposed in \cite{Gabai:2019ryw}, where the authors obtained
bounds on the position of resonances in the complex $s$ plane in the case of Ising Field Theory in $1+1$ dimensions.

Finally, we could explore processes involving particles other than pions. For instance, Glueball couplings influence the lattice energy levels through the Glueball-Glueball phase shift in a non-perturbative manner.
By combining the scattering amplitudes constructed in \cite{Guerrieri:2023qbg} with lattice energy levels, we could gain insights into observables that may otherwise be difficult to measure using standard tools.

\section*{Acknowledgements}

We thank Mattia Bruno, Gilberto Colangelo, Miguel Correia, Barak Gabai, Victor Gorbenko, David Gross, Aditya Hebbar, Denis Karateev, Luciano Maiani, Harish Murali, Jose Ramon Pelaez, Joao Penedones, Alessandro Pilloni, Jiaxin Qiao, Riccardo
Rattazzi, Slava Rychkov, Balt van Rees, David Simmons-Duffin, Pedro Vieira, and Alexander Zhiboedov for useful discussions. 

We thank Gilberto Colangelo, Joao Penedones, Alessandro Pilloni, Balt van Rees, Pedro Vieira, Mark Wise and Alexander Zhiboedov for the many useful comments on the draft.

Research at the Perimeter Institute is supported in part by the Government of
Canada through NSERC and by the Province of Ontario
through MRI.  
AG is supported by the European
Union - NextGenerationEU, under the programme Seal
of Excellence@UNIPD, project acronym CluEs.
The work of KH is supported by the Simons Foundation grant 488649 (Simons Collaboration on the Nonperturbative Bootstrap), by the Swiss National Science Foundation through the project 200020 197160, through the National Centre of Competence in Research SwissMAP and by the Simons Collaboration on Celestial Holography.
NS's work is supported by Simons Foundation grant 488657 (Simons Collaboration on the Nonperturbative Bootstrap).

\appendix

\begin{figure*}[t]
    \centering
    \includegraphics[width=\linewidth]{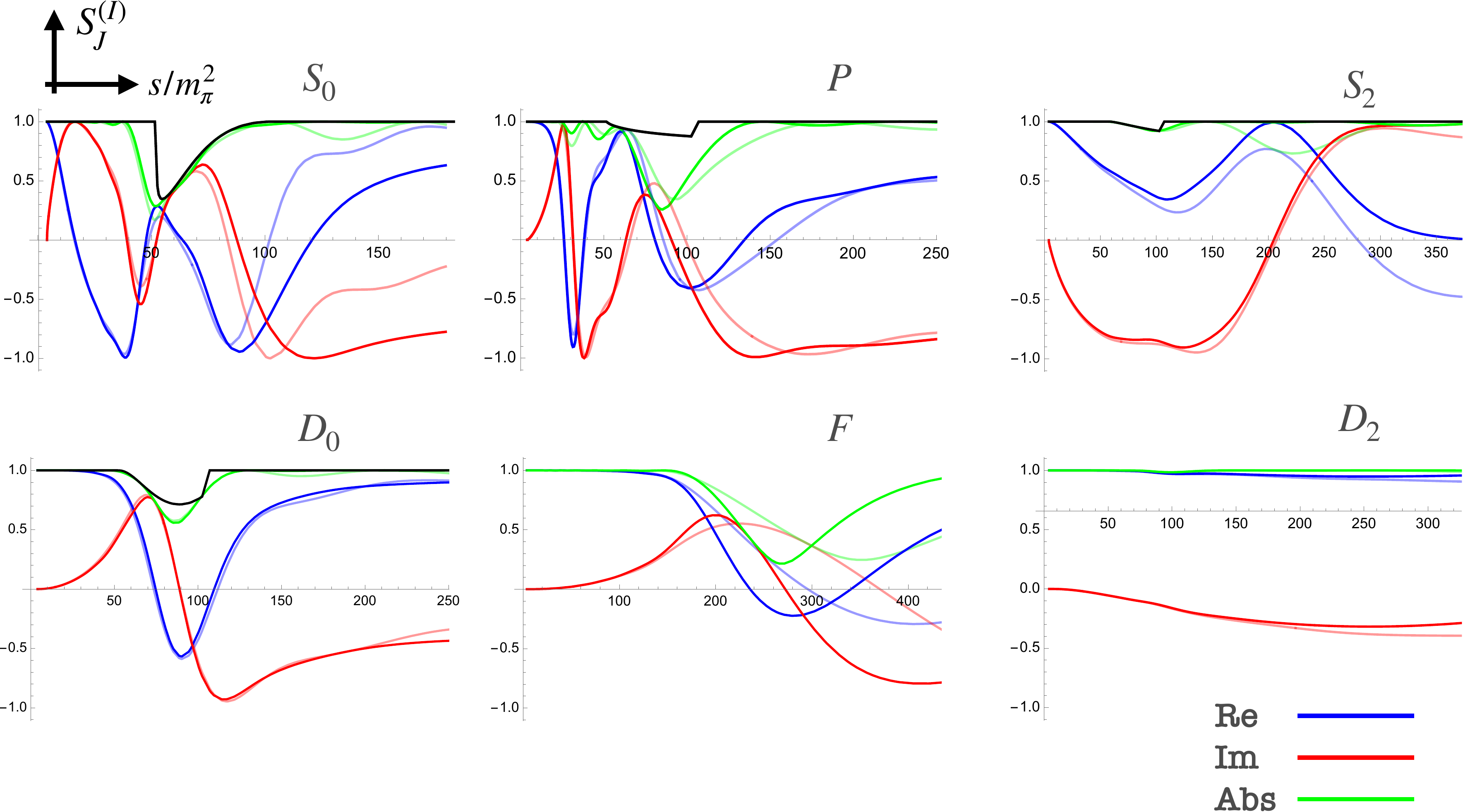}
    \caption{Real (blue), imaginary (red), and absolute value (green) of the phase shifts for various channels. Thick lines are from our best numerics with $(N,M)=(14,12)$, light lines for $(N,M)=(12,10)$. The elasticity profiles used as a constraint are denoted in black. We do not draw it for the channels where we impose the simple probability conservation $|S_J^{(I)}|\leq 1$. }
    \label{fig:raw_data}
\end{figure*}

\section{Details of the S-matrix bootstrap numerics} 
\label{sec:numerical_details}

The ansatz used to produce the results of this letter has multi-foliation  $\Sigma=\{20/3,30,50,86\}$.
As discussed in \cite{Gaikwad:2023hof}, to best approximate resonant structures it is convenient to set the centers of the foliations around their expected positions. That explains our choice for the set $\Sigma$ which is tuned on the expected position of the $\rho(770)$ with $m^2_\rho\approx 30$, the $f_0(980)$ with $m^2_{f_0}\approx 50$, and the $f_2(1260)$ with $m^2_{f_2}\approx 86$.

Let us call $N=N_{20/3}$, and $M=N_\sigma$, when $\sigma$ is any other foliation. To produce the plots in the main text we used $N=14$, and $M=12$ for a total of $N_\text{vars}=547$ variables.
 We have also performed a second run with $N=12$, and $M=10$ for a total of $N_\text{vars}=397$ to check the systematic due to the size of the ansatz.

To impose the unitarity constraints we project numerically our ansatz into partial waves. We use a union of grids, one for each foliation.
The grid is defined through the map 
\beq
s(\phi)=\frac{\sigma-(8-\sigma)\cos\phi}{1+\cos\phi}
\eeq
where $0\leq \phi \leq \pi$ is the upper boundary of the unit disk.
We finally discretize $\phi$ on the roots of the Chebyschev polynomials
\beq
\phi_k=\frac{\pi}{2}\left(1+\cos\frac{\pi k}{N_\text{points}+1}\right),
\eeq
where $k=1,\dots, N_\text{points}$. The maximum number of points used is $N_\text{points}=300$ for $\sigma=20/3$, and $N_\text{points}=150$ for all other foliations. We have performed numerical tests using grids with different numbers of points finding no significant dependence on it.

We have projected our ansatz onto partial waves \eqref{eq:parital_wave_projection} up to spin $\ell=12$, and run our numerics with both maximum spin $L_\text{max}=10,12$. The reason that we can keep $L_\text{max}$ so low is due to the addition of improved positivity constraints \cite{EliasMiro:2022xaa}
\beq
\text{Im}\, \mathcal{T}^{(I)}(s,t)-16\pi\sum_{\ell=0}^{L_\text{max}}(2\ell+1)P_\ell(1+2\tfrac{t}{s-4}) \text{Im}\, t_\ell^{(I)}(s)\geq 0
\eeq
for $0\leq t<4$, $s>4$, and $I=0,1,2$. By imposing this condition we constrain the tail of spins higher than $L_\text{max}$ that we do not explicitly bound with numerical unitarity. 

In figure \ref{fig:raw_data} we show the partial S-matrices $S_{J}^{(I)}$ obtained using two ansatzes respectively with $(N,M)=(14,2)$ (dark lines), $(N,M)=(12,10)$ (light lines). We plot their real (blue), imaginary (red), and absolute values (green). The black solid line is the inelasticity profile due to $\pi\pi\to K\bar K$ taken from \cite{Garcia-Martin:2011iqs} (for the $S_0$ channel we assume the big-dip scenario) that we impose up to $s\approx 100$. Above that energy we keep the inequality $|S_J^{(I)}|\leq 1$.

The unitarity constraints are sufficiently saturated in all these channels except for the region around $s\approx 100$ in the P wave around the position of the $\rho^\prime$, and in the $F$ wave around $s\approx 250$ at the position of the $\rho_3$. These dips in unitarity around resonances are a typical numerical artifact of our finite $N_\text{vars}$ truncation. It would be interesting to increase both $N$ and $M$ to higher values and see if the positions of these particles better align with their experimental determinations.

\begin{figure*}[t]
    \centering
    \includegraphics[width=\linewidth]{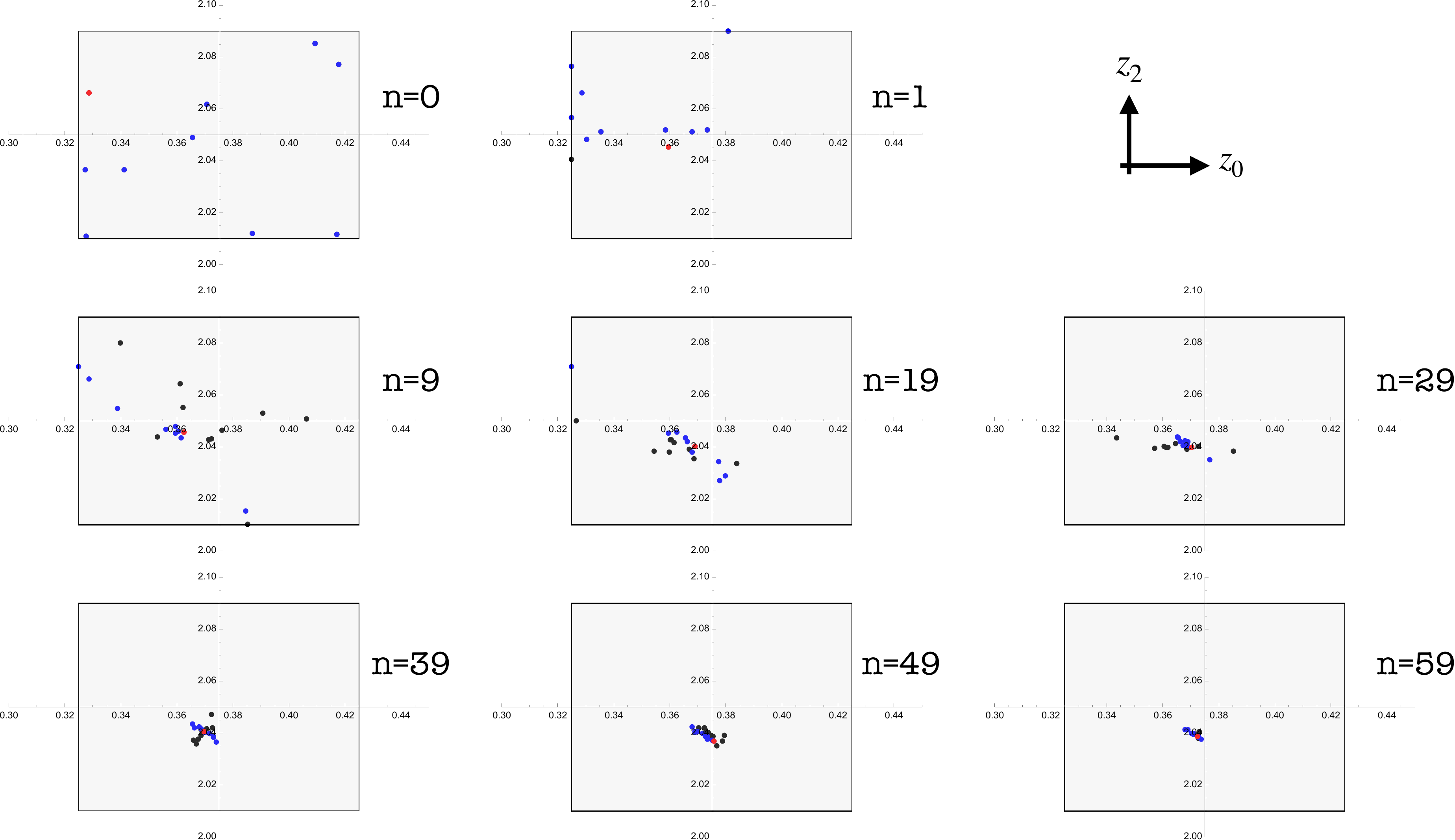}
    \caption{History of the PSO search in the plane $(z_0, z_2)$. At each step $n$, we plot in black the positions of the particles and highlight the best individualistic positions $X_n^{(i)}$ in blue, and the collective best position $Y_n$ in red. As $n$ increases the particles slowly concentrate around the optimal region. The complete movie can be seen in \url{https://giphy.com/gifs/Q5mtDePqPtfSqaQI2z}. }
    \label{fig:swarm}
\end{figure*}

\section{Particle Swarm implementation} 
\label{sec:swarm}

Here we describe our implementation of the PSO algorithm in detail.

To begin the algorithm, we first choose a search region in the fit parameter space where we will look for the minimum of the $\chi^2(\Theta)$ function.
Specifically, we select a region $\vec\Theta_\text{min}\leq \vec \Theta\leq \vec \Theta_\text{max}$ centered around the phenomenological estimates of the fit parameters.

We set the number of particles to $n_\text{p}=10$. This number represents the size of the ``swarm''.
The initial positions of the particles are generated randomly, with values drawn from a Gaussian distribution centered at the region's center, and a variance proportional to its size, denoted by $\vec \epsilon=\tfrac{1}{2}(\vec \Theta_\text{max}-\vec \Theta_\text{min})$.
Similarly, the initial velocities of the particles are sampled from a Gaussian distribution with zero mean and variance $ \vec \epsilon$.

We start the algorithm by updating the positions and velocities of the particles according to equation \eqref{eq:PSO_update}. Additionally, if the particle $i$ moves outside the boundaries of the box during the update step along some direction $j$, we then place it back on the boundary setting $(\Theta_{n+1}^{(i)})_j=(\Theta^{(i)}_{\text{max/min}})_j$ with the opposite velocity.

Evaluating 
$\chi^2(\Theta)$ for a given set of fit parameters is computationally expensive, as it involves solving an S-matrix bootstrap problem. To efficiently estimate the minimum of the 
$\chi^2(\Theta)$ function in fewer steps, we use a modified version of the standard PSO, called \emph{adaptive PSO}.
We use velocity information to adjust the value of $\omega$, as described in \cite{Xu2013AnAP}. 

The goal is to adjust the friction parameter 
$\omega$ using a control function. The value of 
$\omega$ determines how effectively the swarm explores the parameter space. If 
$\omega$ is too large (of order one), the particles will explore erratically, risking failure to converge on the minimum. If $\omega$ is too small, the particles will freeze early, preventing a broad exploration.

In \cite{Xu2013AnAP}, the authors suggest to look at the total velocity of the entire swarm at each step $n$ defined as
\beq
\mathbb{V}_n=\frac{1}{n_\text{p} |\Theta|}\sum_{i=1}^{n_\text{p}} |v_n^{(i)}|,
\eeq
and compare it with a proposed ideal velocity profile
\beq
\mathbb{V}^\text{ideal}_n=\frac{|\epsilon|}{2}\left(1+\cos\left( \frac{\pi n}{0.95 N_\text{iter}}\right)\right).
\eeq

If $\mathbb{V}_{n-1}\geq \mathbb{V}^\text{ideal}_n$, we set $\omega_n=\max(\omega_{n-1}-\Delta\omega,\omega_\text{min})$, otherwise $\omega_n=\min(\omega_{n-1}+\Delta\omega,\omega_\text{max})$. In our case, $\omega_\text{max}=0.8$, $\omega_\text{min}=0.3$, and $\Delta\omega=0.1$. 
We also set $c_1=c_2=1.5$.

The expression of $\mathbb{V}^\text{ideal}$ contains the parameter $N_\text{iter}$ which is the max number of iterations of the search. The ideal velocity is a monotonic decreasing function that starts at $|\epsilon|$ when $n=0$. As we approach the end of the run when $n\sim N_\text{iter}$, the ideal velocity is so small that it forces the friction to decrease to its minimum value $\omega_\text{min}$ hence freezing the search in a neighbor of the global best found up to that point.

The risk with this approach is that choosing a small $N_\text{iter}$ could lead to getting stuck in a local minimum. To mitigate this, we performed two searches with different values of  $N_\text{iter}=30$ and 60.

Figure \ref{fig:swarm} shows the projection of particle positions onto the $(z_0,z_2)$ plane at various steps $n$ during a run with $N_\text{iter}=60$.
Black dots represent the particles' current positions $\Theta_n^{(i)}$ at step $n$, 
blue dots indicate each particle's best position $X_n^{(i)}$ up to that step, and the red dot marks the global best position $Y_n$ found by the swarm.
At $n=0$, the black and blue dots coincide. The gray region represents the search domain.

\newpage

\bibliographystyle{apsrev4-1}
\bibliography{dual_bootstrap}

\begin{thebibliography}{76}%
\makeatletter
\providecommand \@ifxundefined [1]{%
 \@ifx{#1\undefined}
}%
\providecommand \@ifnum [1]{%
 \ifnum #1\expandafter \@firstoftwo
 \else \expandafter \@secondoftwo
 \fi
}%
\providecommand \@ifx [1]{%
 \ifx #1\expandafter \@firstoftwo
 \else \expandafter \@secondoftwo
 \fi
}%
\providecommand \natexlab [1]{#1}%
\providecommand \enquote  [1]{``#1''}%
\providecommand \bibnamefont  [1]{#1}%
\providecommand \bibfnamefont [1]{#1}%
\providecommand \citenamefont [1]{#1}%
\providecommand \href@noop [0]{\@secondoftwo}%
\providecommand \href [0]{\begingroup \@sanitize@url \@href}%
\providecommand \@href[1]{\@@startlink{#1}\@@href}%
\providecommand \@@href[1]{\endgroup#1\@@endlink}%
\providecommand \@sanitize@url [0]{\catcode `\\12\catcode `\$12\catcode `\&12\catcode `\#12\catcode `\^12\catcode `\_12\catcode `\%12\relax}%
\providecommand \@@startlink[1]{}%
\providecommand \@@endlink[0]{}%
\providecommand \url  [0]{\begingroup\@sanitize@url \@url }%
\providecommand \@url [1]{\endgroup\@href {#1}{\urlprefix }}%
\providecommand \urlprefix  [0]{URL }%
\providecommand \Eprint [0]{\href }%
\providecommand \doibase [0]{http://dx.doi.org/}%
\providecommand \selectlanguage [0]{\@gobble}%
\providecommand \bibinfo  [0]{\@secondoftwo}%
\providecommand \bibfield  [0]{\@secondoftwo}%
\providecommand \translation [1]{[#1]}%
\providecommand \BibitemOpen [0]{}%
\providecommand \bibitemStop [0]{}%
\providecommand \bibitemNoStop [0]{.\EOS\space}%
\providecommand \EOS [0]{\spacefactor3000\relax}%
\providecommand \BibitemShut  [1]{\csname bibitem#1\endcsname}%
\let\auto@bib@innerbib\@empty
\bibitem [{\citenamefont {Roy}(1971)}]{Roy:1971tc}%
  \BibitemOpen
  \bibfield  {author} {\bibinfo {author} {\bibfnamefont {S.~M.}\ \bibnamefont {Roy}},\ }\href {\doibase 10.1016/0370-2693(71)90724-6} {\bibfield  {journal} {\bibinfo  {journal} {Phys. Lett. B}\ }\textbf {\bibinfo {volume} {36}},\ \bibinfo {pages} {353} (\bibinfo {year} {1971})}\BibitemShut {NoStop}%
\bibitem [{\citenamefont {Ananthanarayan}\ \emph {et~al.}(2001)\citenamefont {Ananthanarayan}, \citenamefont {Colangelo}, \citenamefont {Gasser},\ and\ \citenamefont {Leutwyler}}]{Ananthanarayan:2000ht}%
  \BibitemOpen
  \bibfield  {author} {\bibinfo {author} {\bibfnamefont {B.}~\bibnamefont {Ananthanarayan}}, \bibinfo {author} {\bibfnamefont {G.}~\bibnamefont {Colangelo}}, \bibinfo {author} {\bibfnamefont {J.}~\bibnamefont {Gasser}}, \ and\ \bibinfo {author} {\bibfnamefont {H.}~\bibnamefont {Leutwyler}},\ }\href {\doibase 10.1016/S0370-1573(01)00009-6} {\bibfield  {journal} {\bibinfo  {journal} {Phys. Rept.}\ }\textbf {\bibinfo {volume} {353}},\ \bibinfo {pages} {207} (\bibinfo {year} {2001})},\ \Eprint {http://arxiv.org/abs/hep-ph/0005297} {arXiv:hep-ph/0005297} \BibitemShut {NoStop}%
\bibitem [{\citenamefont {Guerrieri}\ \emph {et~al.}(2019)\citenamefont {Guerrieri}, \citenamefont {Penedones},\ and\ \citenamefont {Vieira}}]{Andrea}%
  \BibitemOpen
  \bibfield  {author} {\bibinfo {author} {\bibfnamefont {A.~L.}\ \bibnamefont {Guerrieri}}, \bibinfo {author} {\bibfnamefont {J.}~\bibnamefont {Penedones}}, \ and\ \bibinfo {author} {\bibfnamefont {P.}~\bibnamefont {Vieira}},\ }\href {\doibase 10.1103/PhysRevLett.122.241604} {\bibfield  {journal} {\bibinfo  {journal} {Phys. Rev. Lett.}\ }\textbf {\bibinfo {volume} {122}},\ \bibinfo {pages} {241604} (\bibinfo {year} {2019})},\ \Eprint {http://arxiv.org/abs/1810.12849} {arXiv:1810.12849 [hep-th]} \BibitemShut {NoStop}%
\bibitem [{\citenamefont {Guerrieri}\ \emph {et~al.}(2021{\natexlab{a}})\citenamefont {Guerrieri}, \citenamefont {Penedones},\ and\ \citenamefont {Vieira}}]{Guerrieri:2020bto}%
  \BibitemOpen
  \bibfield  {author} {\bibinfo {author} {\bibfnamefont {A.~L.}\ \bibnamefont {Guerrieri}}, \bibinfo {author} {\bibfnamefont {J.}~\bibnamefont {Penedones}}, \ and\ \bibinfo {author} {\bibfnamefont {P.}~\bibnamefont {Vieira}},\ }\href {\doibase 10.1007/JHEP06(2021)088} {\bibfield  {journal} {\bibinfo  {journal} {JHEP}\ }\textbf {\bibinfo {volume} {06}},\ \bibinfo {pages} {088} (\bibinfo {year} {2021}{\natexlab{a}})},\ \Eprint {http://arxiv.org/abs/2011.02802} {arXiv:2011.02802 [hep-th]} \BibitemShut {NoStop}%
\bibitem [{\citenamefont {Bose}\ \emph {et~al.}(2020)\citenamefont {Bose}, \citenamefont {Haldar}, \citenamefont {Sinha}, \citenamefont {Sinha},\ and\ \citenamefont {Tiwari}}]{Bose:2020shm}%
  \BibitemOpen
  \bibfield  {author} {\bibinfo {author} {\bibfnamefont {A.}~\bibnamefont {Bose}}, \bibinfo {author} {\bibfnamefont {P.}~\bibnamefont {Haldar}}, \bibinfo {author} {\bibfnamefont {A.}~\bibnamefont {Sinha}}, \bibinfo {author} {\bibfnamefont {P.}~\bibnamefont {Sinha}}, \ and\ \bibinfo {author} {\bibfnamefont {S.~S.}\ \bibnamefont {Tiwari}},\ }\href {\doibase 10.21468/SciPostPhys.9.5.081} {\bibfield  {journal} {\bibinfo  {journal} {SciPost Phys.}\ }\textbf {\bibinfo {volume} {9}},\ \bibinfo {pages} {081} (\bibinfo {year} {2020})},\ \Eprint {http://arxiv.org/abs/2006.12213} {arXiv:2006.12213 [hep-th]} \BibitemShut {NoStop}%
\bibitem [{\citenamefont {Bose}\ \emph {et~al.}(2021)\citenamefont {Bose}, \citenamefont {Sinha},\ and\ \citenamefont {Tiwari}}]{Bose:2020cod}%
  \BibitemOpen
  \bibfield  {author} {\bibinfo {author} {\bibfnamefont {A.}~\bibnamefont {Bose}}, \bibinfo {author} {\bibfnamefont {A.}~\bibnamefont {Sinha}}, \ and\ \bibinfo {author} {\bibfnamefont {S.~S.}\ \bibnamefont {Tiwari}},\ }\href {\doibase 10.21468/SciPostPhys.10.5.122} {\bibfield  {journal} {\bibinfo  {journal} {SciPost Phys.}\ }\textbf {\bibinfo {volume} {10}},\ \bibinfo {pages} {122} (\bibinfo {year} {2021})},\ \Eprint {http://arxiv.org/abs/2011.07944} {arXiv:2011.07944 [hep-th]} \BibitemShut {NoStop}%
\bibitem [{\citenamefont {He}\ and\ \citenamefont {Kruczenski}(2023)}]{He:2023lyy}%
  \BibitemOpen
  \bibfield  {author} {\bibinfo {author} {\bibfnamefont {Y.}~\bibnamefont {He}}\ and\ \bibinfo {author} {\bibfnamefont {M.}~\bibnamefont {Kruczenski}},\ }\href@noop {} {\  (\bibinfo {year} {2023})},\ \Eprint {http://arxiv.org/abs/2309.12402} {arXiv:2309.12402 [hep-th]} \BibitemShut {NoStop}%
\bibitem [{\citenamefont {He}\ and\ \citenamefont {Kruczenski}(2024)}]{He:2024nwd}%
  \BibitemOpen
  \bibfield  {author} {\bibinfo {author} {\bibfnamefont {Y.}~\bibnamefont {He}}\ and\ \bibinfo {author} {\bibfnamefont {M.}~\bibnamefont {Kruczenski}},\ }\href@noop {} {\  (\bibinfo {year} {2024})},\ \Eprint {http://arxiv.org/abs/2403.10772} {arXiv:2403.10772 [hep-th]} \BibitemShut {NoStop}%
\bibitem [{\citenamefont {Albert}\ and\ \citenamefont {Rastelli}(2022)}]{Albert:2022oes}%
  \BibitemOpen
  \bibfield  {author} {\bibinfo {author} {\bibfnamefont {J.}~\bibnamefont {Albert}}\ and\ \bibinfo {author} {\bibfnamefont {L.}~\bibnamefont {Rastelli}},\ }\href {\doibase 10.1007/JHEP08(2022)151} {\bibfield  {journal} {\bibinfo  {journal} {JHEP}\ }\textbf {\bibinfo {volume} {08}},\ \bibinfo {pages} {151} (\bibinfo {year} {2022})},\ \Eprint {http://arxiv.org/abs/2203.11950} {arXiv:2203.11950 [hep-th]} \BibitemShut {NoStop}%
\bibitem [{\citenamefont {Fernandez}\ \emph {et~al.}(2023)\citenamefont {Fernandez}, \citenamefont {Pomarol}, \citenamefont {Riva},\ and\ \citenamefont {Sciotti}}]{Fernandez:2022kzi}%
  \BibitemOpen
  \bibfield  {author} {\bibinfo {author} {\bibfnamefont {C.}~\bibnamefont {Fernandez}}, \bibinfo {author} {\bibfnamefont {A.}~\bibnamefont {Pomarol}}, \bibinfo {author} {\bibfnamefont {F.}~\bibnamefont {Riva}}, \ and\ \bibinfo {author} {\bibfnamefont {F.}~\bibnamefont {Sciotti}},\ }\href {\doibase 10.1007/JHEP06(2023)094} {\bibfield  {journal} {\bibinfo  {journal} {JHEP}\ }\textbf {\bibinfo {volume} {06}},\ \bibinfo {pages} {094} (\bibinfo {year} {2023})},\ \Eprint {http://arxiv.org/abs/2211.12488} {arXiv:2211.12488 [hep-th]} \BibitemShut {NoStop}%
\bibitem [{\citenamefont {Albert}\ and\ \citenamefont {Rastelli}(2024)}]{Albert:2023jtd}%
  \BibitemOpen
  \bibfield  {author} {\bibinfo {author} {\bibfnamefont {J.}~\bibnamefont {Albert}}\ and\ \bibinfo {author} {\bibfnamefont {L.}~\bibnamefont {Rastelli}},\ }\href {\doibase 10.1007/JHEP09(2024)039} {\bibfield  {journal} {\bibinfo  {journal} {JHEP}\ }\textbf {\bibinfo {volume} {09}},\ \bibinfo {pages} {039} (\bibinfo {year} {2024})},\ \Eprint {http://arxiv.org/abs/2307.01246} {arXiv:2307.01246 [hep-th]} \BibitemShut {NoStop}%
\bibitem [{\citenamefont {Albert}\ \emph {et~al.}(2024)\citenamefont {Albert}, \citenamefont {Henriksson}, \citenamefont {Rastelli},\ and\ \citenamefont {Vichi}}]{Albert:2023seb}%
  \BibitemOpen
  \bibfield  {author} {\bibinfo {author} {\bibfnamefont {J.}~\bibnamefont {Albert}}, \bibinfo {author} {\bibfnamefont {J.}~\bibnamefont {Henriksson}}, \bibinfo {author} {\bibfnamefont {L.}~\bibnamefont {Rastelli}}, \ and\ \bibinfo {author} {\bibfnamefont {A.}~\bibnamefont {Vichi}},\ }\href {\doibase 10.1007/JHEP09(2024)172} {\bibfield  {journal} {\bibinfo  {journal} {JHEP}\ }\textbf {\bibinfo {volume} {09}},\ \bibinfo {pages} {172} (\bibinfo {year} {2024})},\ \Eprint {http://arxiv.org/abs/2312.15013} {arXiv:2312.15013 [hep-th]} \BibitemShut {NoStop}%
\bibitem [{\citenamefont {Navas}\ \emph {et~al.}(2024)\citenamefont {Navas} \emph {et~al.}}]{ParticleDataGroup:2024cfk}%
  \BibitemOpen
  \bibfield  {author} {\bibinfo {author} {\bibfnamefont {S.}~\bibnamefont {Navas}} \emph {et~al.} (\bibinfo {collaboration} {Particle Data Group}),\ }\href {\doibase 10.1103/PhysRevD.110.030001} {\bibfield  {journal} {\bibinfo  {journal} {Phys. Rev. D}\ }\textbf {\bibinfo {volume} {110}},\ \bibinfo {pages} {030001} (\bibinfo {year} {2024})}\BibitemShut {NoStop}%
\bibitem [{\citenamefont {Jaffe}(1977)}]{Jaffe:1976ig}%
  \BibitemOpen
  \bibfield  {author} {\bibinfo {author} {\bibfnamefont {R.~L.}\ \bibnamefont {Jaffe}},\ }\href {\doibase 10.1103/PhysRevD.15.267} {\bibfield  {journal} {\bibinfo  {journal} {Phys. Rev. D}\ }\textbf {\bibinfo {volume} {15}},\ \bibinfo {pages} {267} (\bibinfo {year} {1977})}\BibitemShut {NoStop}%
\bibitem [{\citenamefont {'t~Hooft}\ \emph {et~al.}(2008)\citenamefont {'t~Hooft}, \citenamefont {Isidori}, \citenamefont {Maiani}, \citenamefont {Polosa},\ and\ \citenamefont {Riquer}}]{tHooft:2008rus}%
  \BibitemOpen
  \bibfield  {author} {\bibinfo {author} {\bibfnamefont {G.}~\bibnamefont {'t~Hooft}}, \bibinfo {author} {\bibfnamefont {G.}~\bibnamefont {Isidori}}, \bibinfo {author} {\bibfnamefont {L.}~\bibnamefont {Maiani}}, \bibinfo {author} {\bibfnamefont {A.~D.}\ \bibnamefont {Polosa}}, \ and\ \bibinfo {author} {\bibfnamefont {V.}~\bibnamefont {Riquer}},\ }\href {\doibase 10.1016/j.physletb.2008.03.036} {\bibfield  {journal} {\bibinfo  {journal} {Phys. Lett. B}\ }\textbf {\bibinfo {volume} {662}},\ \bibinfo {pages} {424} (\bibinfo {year} {2008})},\ \Eprint {http://arxiv.org/abs/0801.2288} {arXiv:0801.2288 [hep-ph]} \BibitemShut {NoStop}%
\bibitem [{\citenamefont {Paulos}\ \emph {et~al.}(2019)\citenamefont {Paulos}, \citenamefont {Penedones}, \citenamefont {Toledo}, \citenamefont {van Rees},\ and\ \citenamefont {Vieira}}]{Paper3}%
  \BibitemOpen
  \bibfield  {author} {\bibinfo {author} {\bibfnamefont {M.~F.}\ \bibnamefont {Paulos}}, \bibinfo {author} {\bibfnamefont {J.}~\bibnamefont {Penedones}}, \bibinfo {author} {\bibfnamefont {J.}~\bibnamefont {Toledo}}, \bibinfo {author} {\bibfnamefont {B.~C.}\ \bibnamefont {van Rees}}, \ and\ \bibinfo {author} {\bibfnamefont {P.}~\bibnamefont {Vieira}},\ }\href {\doibase 10.1007/JHEP12(2019)040} {\bibfield  {journal} {\bibinfo  {journal} {JHEP}\ }\textbf {\bibinfo {volume} {12}},\ \bibinfo {pages} {040} (\bibinfo {year} {2019})},\ \Eprint {http://arxiv.org/abs/1708.06765} {arXiv:1708.06765 [hep-th]} \BibitemShut {NoStop}%
\bibitem [{\citenamefont {Guerrieri}\ and\ \citenamefont {Sever}(2021)}]{Guerrieri:2021tak}%
  \BibitemOpen
  \bibfield  {author} {\bibinfo {author} {\bibfnamefont {A.}~\bibnamefont {Guerrieri}}\ and\ \bibinfo {author} {\bibfnamefont {A.}~\bibnamefont {Sever}},\ }\href {\doibase 10.1103/PhysRevLett.127.251601} {\bibfield  {journal} {\bibinfo  {journal} {Phys. Rev. Lett.}\ }\textbf {\bibinfo {volume} {127}},\ \bibinfo {pages} {251601} (\bibinfo {year} {2021})},\ \Eprint {http://arxiv.org/abs/2106.10257} {arXiv:2106.10257 [hep-th]} \BibitemShut {NoStop}%
\bibitem [{\citenamefont {Elias~Miro}\ \emph {et~al.}(2024)\citenamefont {Elias~Miro}, \citenamefont {Guerrieri},\ and\ \citenamefont {Gumus}}]{EliasMiro:2023fqi}%
  \BibitemOpen
  \bibfield  {author} {\bibinfo {author} {\bibfnamefont {J.}~\bibnamefont {Elias~Miro}}, \bibinfo {author} {\bibfnamefont {A.~L.}\ \bibnamefont {Guerrieri}}, \ and\ \bibinfo {author} {\bibfnamefont {M.~A.}\ \bibnamefont {Gumus}},\ }\href {\doibase 10.1103/PhysRevD.110.016007} {\bibfield  {journal} {\bibinfo  {journal} {Phys. Rev. D}\ }\textbf {\bibinfo {volume} {110}},\ \bibinfo {pages} {016007} (\bibinfo {year} {2024})},\ \Eprint {http://arxiv.org/abs/2311.09283} {arXiv:2311.09283 [hep-ph]} \BibitemShut {NoStop}%
\bibitem [{\citenamefont {Guerrieri}\ \emph {et~al.}(2023{\natexlab{a}})\citenamefont {Guerrieri}, \citenamefont {Hebbar},\ and\ \citenamefont {van Rees}}]{Guerrieri:2023qbg}%
  \BibitemOpen
  \bibfield  {author} {\bibinfo {author} {\bibfnamefont {A.~L.}\ \bibnamefont {Guerrieri}}, \bibinfo {author} {\bibfnamefont {A.}~\bibnamefont {Hebbar}}, \ and\ \bibinfo {author} {\bibfnamefont {B.~C.}\ \bibnamefont {van Rees}},\ }\href@noop {} {\  (\bibinfo {year} {2023}{\natexlab{a}})},\ \Eprint {http://arxiv.org/abs/2312.00127} {arXiv:2312.00127 [hep-th]} \BibitemShut {NoStop}%
\bibitem [{\citenamefont {Martin}(1966)}]{Martin2}%
  \BibitemOpen
  \bibfield  {author} {\bibinfo {author} {\bibfnamefont {A.}~\bibnamefont {Martin}},\ }\href {\doibase 10.1007/BF02719361} {\bibfield  {journal} {\bibinfo  {journal} {Nuovo Cim. A}\ }\textbf {\bibinfo {volume} {44}},\ \bibinfo {pages} {1219} (\bibinfo {year} {1966})}\BibitemShut {NoStop}%
\bibitem [{\citenamefont {Elias~Miro}\ \emph {et~al.}(2023)\citenamefont {Elias~Miro}, \citenamefont {Guerrieri},\ and\ \citenamefont {Gumus}}]{EliasMiro:2022xaa}%
  \BibitemOpen
  \bibfield  {author} {\bibinfo {author} {\bibfnamefont {J.}~\bibnamefont {Elias~Miro}}, \bibinfo {author} {\bibfnamefont {A.}~\bibnamefont {Guerrieri}}, \ and\ \bibinfo {author} {\bibfnamefont {M.~A.}\ \bibnamefont {Gumus}},\ }\href {\doibase 10.1007/JHEP05(2023)001} {\bibfield  {journal} {\bibinfo  {journal} {JHEP}\ }\textbf {\bibinfo {volume} {05}},\ \bibinfo {pages} {001} (\bibinfo {year} {2023})},\ \Eprint {http://arxiv.org/abs/2210.01502} {arXiv:2210.01502 [hep-th]} \BibitemShut {NoStop}%
\bibitem [{\citenamefont {Gaikwad}\ \emph {et~al.}(2024)\citenamefont {Gaikwad}, \citenamefont {Gorbenko},\ and\ \citenamefont {Guerrieri}}]{Gaikwad:2023hof}%
  \BibitemOpen
  \bibfield  {author} {\bibinfo {author} {\bibfnamefont {A.}~\bibnamefont {Gaikwad}}, \bibinfo {author} {\bibfnamefont {V.}~\bibnamefont {Gorbenko}}, \ and\ \bibinfo {author} {\bibfnamefont {A.~L.}\ \bibnamefont {Guerrieri}},\ }\href {\doibase 10.1007/JHEP01(2024)090} {\bibfield  {journal} {\bibinfo  {journal} {JHEP}\ }\textbf {\bibinfo {volume} {01}},\ \bibinfo {pages} {090} (\bibinfo {year} {2024})},\ \Eprint {http://arxiv.org/abs/2310.20698} {arXiv:2310.20698 [hep-th]} \BibitemShut {NoStop}%
\bibitem [{\citenamefont {Antunes}\ \emph {et~al.}(2023)\citenamefont {Antunes}, \citenamefont {Costa},\ and\ \citenamefont {Pereira}}]{Antunes:2023irg}%
  \BibitemOpen
  \bibfield  {author} {\bibinfo {author} {\bibfnamefont {A.}~\bibnamefont {Antunes}}, \bibinfo {author} {\bibfnamefont {M.~S.}\ \bibnamefont {Costa}}, \ and\ \bibinfo {author} {\bibfnamefont {J.}~\bibnamefont {Pereira}},\ }\href {\doibase 10.1016/j.physletb.2023.138225} {\bibfield  {journal} {\bibinfo  {journal} {Phys. Lett. B}\ }\textbf {\bibinfo {volume} {846}},\ \bibinfo {pages} {138225} (\bibinfo {year} {2023})},\ \Eprint {http://arxiv.org/abs/2301.13219} {arXiv:2301.13219 [hep-th]} \BibitemShut {NoStop}%
\bibitem [{\citenamefont {Guerrieri}\ \emph {et~al.}(2024)\citenamefont {Guerrieri}, \citenamefont {Homrich},\ and\ \citenamefont {Vieira}}]{Guerrieri:2024ckc}%
  \BibitemOpen
  \bibfield  {author} {\bibinfo {author} {\bibfnamefont {A.}~\bibnamefont {Guerrieri}}, \bibinfo {author} {\bibfnamefont {A.}~\bibnamefont {Homrich}}, \ and\ \bibinfo {author} {\bibfnamefont {P.}~\bibnamefont {Vieira}},\ }\href@noop {} {\  (\bibinfo {year} {2024})},\ \Eprint {http://arxiv.org/abs/2404.10812} {arXiv:2404.10812 [hep-th]} \BibitemShut {NoStop}%
\bibitem [{\citenamefont {Tourkine}\ and\ \citenamefont {Zhiboedov}(2023)}]{Tourkine:2023xtu}%
  \BibitemOpen
  \bibfield  {author} {\bibinfo {author} {\bibfnamefont {P.}~\bibnamefont {Tourkine}}\ and\ \bibinfo {author} {\bibfnamefont {A.}~\bibnamefont {Zhiboedov}},\ }\href {\doibase 10.1007/JHEP11(2023)005} {\bibfield  {journal} {\bibinfo  {journal} {JHEP}\ }\textbf {\bibinfo {volume} {11}},\ \bibinfo {pages} {005} (\bibinfo {year} {2023})},\ \Eprint {http://arxiv.org/abs/2303.08839} {arXiv:2303.08839 [hep-th]} \BibitemShut {NoStop}%
\bibitem [{\citenamefont {Garcia-Martin}\ \emph {et~al.}(2011)\citenamefont {Garcia-Martin}, \citenamefont {Kaminski}, \citenamefont {Pelaez}, \citenamefont {Ruiz~de Elvira},\ and\ \citenamefont {Yndurain}}]{Garcia-Martin:2011iqs}%
  \BibitemOpen
  \bibfield  {author} {\bibinfo {author} {\bibfnamefont {R.}~\bibnamefont {Garcia-Martin}}, \bibinfo {author} {\bibfnamefont {R.}~\bibnamefont {Kaminski}}, \bibinfo {author} {\bibfnamefont {J.~R.}\ \bibnamefont {Pelaez}}, \bibinfo {author} {\bibfnamefont {J.}~\bibnamefont {Ruiz~de Elvira}}, \ and\ \bibinfo {author} {\bibfnamefont {F.~J.}\ \bibnamefont {Yndurain}},\ }\href {\doibase 10.1103/PhysRevD.83.074004} {\bibfield  {journal} {\bibinfo  {journal} {Phys. Rev. D}\ }\textbf {\bibinfo {volume} {83}},\ \bibinfo {pages} {074004} (\bibinfo {year} {2011})},\ \Eprint {http://arxiv.org/abs/1102.2183} {arXiv:1102.2183 [hep-ph]} \BibitemShut {NoStop}%
\bibitem [{\citenamefont {Homrich}\ \emph {et~al.}(2019)\citenamefont {Homrich}, \citenamefont {Penedones}, \citenamefont {Toledo}, \citenamefont {van Rees},\ and\ \citenamefont {Vieira}}]{Homrich:2019cbt}%
  \BibitemOpen
  \bibfield  {author} {\bibinfo {author} {\bibfnamefont {A.}~\bibnamefont {Homrich}}, \bibinfo {author} {\bibfnamefont {J.~a.}\ \bibnamefont {Penedones}}, \bibinfo {author} {\bibfnamefont {J.}~\bibnamefont {Toledo}}, \bibinfo {author} {\bibfnamefont {B.~C.}\ \bibnamefont {van Rees}}, \ and\ \bibinfo {author} {\bibfnamefont {P.}~\bibnamefont {Vieira}},\ }\href {\doibase 10.1007/JHEP11(2019)076} {\bibfield  {journal} {\bibinfo  {journal} {JHEP}\ }\textbf {\bibinfo {volume} {11}},\ \bibinfo {pages} {076} (\bibinfo {year} {2019})},\ \Eprint {http://arxiv.org/abs/1905.06905} {arXiv:1905.06905 [hep-th]} \BibitemShut {NoStop}%
\bibitem [{\citenamefont {Bercini}\ \emph {et~al.}(2020)\citenamefont {Bercini}, \citenamefont {Fabri}, \citenamefont {Homrich},\ and\ \citenamefont {Vieira}}]{Bercini:2019vme}%
  \BibitemOpen
  \bibfield  {author} {\bibinfo {author} {\bibfnamefont {C.}~\bibnamefont {Bercini}}, \bibinfo {author} {\bibfnamefont {M.}~\bibnamefont {Fabri}}, \bibinfo {author} {\bibfnamefont {A.}~\bibnamefont {Homrich}}, \ and\ \bibinfo {author} {\bibfnamefont {P.}~\bibnamefont {Vieira}},\ }\href {\doibase 10.1103/PhysRevD.101.045022} {\bibfield  {journal} {\bibinfo  {journal} {Phys. Rev. D}\ }\textbf {\bibinfo {volume} {101}},\ \bibinfo {pages} {045022} (\bibinfo {year} {2020})},\ \Eprint {http://arxiv.org/abs/1909.06453} {arXiv:1909.06453 [hep-th]} \BibitemShut {NoStop}%
\bibitem [{\citenamefont {Guerrieri}\ \emph {et~al.}(2020)\citenamefont {Guerrieri}, \citenamefont {Homrich},\ and\ \citenamefont {Vieira}}]{Guerrieri:2020kcs}%
  \BibitemOpen
  \bibfield  {author} {\bibinfo {author} {\bibfnamefont {A.~L.}\ \bibnamefont {Guerrieri}}, \bibinfo {author} {\bibfnamefont {A.}~\bibnamefont {Homrich}}, \ and\ \bibinfo {author} {\bibfnamefont {P.}~\bibnamefont {Vieira}},\ }\href {\doibase 10.1007/JHEP11(2020)084} {\bibfield  {journal} {\bibinfo  {journal} {JHEP}\ }\textbf {\bibinfo {volume} {11}},\ \bibinfo {pages} {084} (\bibinfo {year} {2020})},\ \Eprint {http://arxiv.org/abs/2008.02770} {arXiv:2008.02770 [hep-th]} \BibitemShut {NoStop}%
\bibitem [{\citenamefont {Adler}(1965)}]{Adler:1964um}%
  \BibitemOpen
  \bibfield  {author} {\bibinfo {author} {\bibfnamefont {S.~L.}\ \bibnamefont {Adler}},\ }\href {\doibase 10.1103/PhysRev.137.B1022} {\bibfield  {journal} {\bibinfo  {journal} {Phys. Rev.}\ }\textbf {\bibinfo {volume} {137}},\ \bibinfo {pages} {B1022} (\bibinfo {year} {1965})}\BibitemShut {NoStop}%
\bibitem [{\citenamefont {Weinberg}(1979)}]{Weinberg:1978kz}%
  \BibitemOpen
  \bibfield  {author} {\bibinfo {author} {\bibfnamefont {S.}~\bibnamefont {Weinberg}},\ }\href {\doibase 10.1016/0378-4371(79)90223-1} {\bibfield  {journal} {\bibinfo  {journal} {Physica A}\ }\textbf {\bibinfo {volume} {96}},\ \bibinfo {pages} {327} (\bibinfo {year} {1979})}\BibitemShut {NoStop}%
\bibitem [{\citenamefont {C\'ordova}\ \emph {et~al.}(2020)\citenamefont {C\'ordova}, \citenamefont {He}, \citenamefont {Kruczenski},\ and\ \citenamefont {Vieira}}]{Cordova:2019lot}%
  \BibitemOpen
  \bibfield  {author} {\bibinfo {author} {\bibfnamefont {L.}~\bibnamefont {C\'ordova}}, \bibinfo {author} {\bibfnamefont {Y.}~\bibnamefont {He}}, \bibinfo {author} {\bibfnamefont {M.}~\bibnamefont {Kruczenski}}, \ and\ \bibinfo {author} {\bibfnamefont {P.}~\bibnamefont {Vieira}},\ }\href {\doibase 10.1007/JHEP04(2020)142} {\bibfield  {journal} {\bibinfo  {journal} {JHEP}\ }\textbf {\bibinfo {volume} {04}},\ \bibinfo {pages} {142} (\bibinfo {year} {2020})},\ \Eprint {http://arxiv.org/abs/1909.06495} {arXiv:1909.06495 [hep-th]} \BibitemShut {NoStop}%
\bibitem [{\citenamefont {Simmons-Duffin}(2015)}]{Simmons-Duffin:2015qma}%
  \BibitemOpen
  \bibfield  {author} {\bibinfo {author} {\bibfnamefont {D.}~\bibnamefont {Simmons-Duffin}},\ }\href {\doibase 10.1007/JHEP06(2015)174} {\bibfield  {journal} {\bibinfo  {journal} {JHEP}\ }\textbf {\bibinfo {volume} {06}},\ \bibinfo {pages} {174} (\bibinfo {year} {2015})},\ \Eprint {http://arxiv.org/abs/1502.02033} {arXiv:1502.02033 [hep-th]} \BibitemShut {NoStop}%
\bibitem [{\citenamefont {Landry}\ and\ \citenamefont {Simmons-Duffin}(2019)}]{Landry:2019qug}%
  \BibitemOpen
  \bibfield  {author} {\bibinfo {author} {\bibfnamefont {W.}~\bibnamefont {Landry}}\ and\ \bibinfo {author} {\bibfnamefont {D.}~\bibnamefont {Simmons-Duffin}},\ }\href@noop {} {\  (\bibinfo {year} {2019})},\ \Eprint {http://arxiv.org/abs/1909.09745} {arXiv:1909.09745 [hep-th]} \BibitemShut {NoStop}%
\bibitem [{\citenamefont {Shi}\ and\ \citenamefont {Eberhart}(1998)}]{Shi1998AMP}%
  \BibitemOpen
  \bibfield  {author} {\bibinfo {author} {\bibfnamefont {Y.}~\bibnamefont {Shi}}\ and\ \bibinfo {author} {\bibfnamefont {R.~C.}\ \bibnamefont {Eberhart}},\ }\href {https://api.semanticscholar.org/CorpusID:16708577} {\bibfield  {journal} {\bibinfo  {journal} {1998 IEEE International Conference on Evolutionary Computation Proceedings. IEEE World Congress on Computational Intelligence (Cat. No.98TH8360)}\ ,\ \bibinfo {pages} {69}} (\bibinfo {year} {1998})}\BibitemShut {NoStop}%
\bibitem [{\citenamefont {Clerc}(2006)}]{PSOclerc}%
  \BibitemOpen
  \bibfield  {author} {\bibinfo {author} {\bibfnamefont {M.}~\bibnamefont {Clerc}},\ }\href@noop {} {\emph {\bibinfo {title} {Particle Swarm Optimization}}}\ (\bibinfo  {publisher} {ISTE Ltd},\ \bibinfo {address} {London, UK},\ \bibinfo {year} {2006})\BibitemShut {NoStop}%
\bibitem [{\citenamefont {Reehorst}\ \emph {et~al.}(2021)\citenamefont {Reehorst}, \citenamefont {Rychkov}, \citenamefont {Simmons-Duffin}, \citenamefont {Sirois}, \citenamefont {Su},\ and\ \citenamefont {van Rees}}]{Reehorst:2021ykw}%
  \BibitemOpen
  \bibfield  {author} {\bibinfo {author} {\bibfnamefont {M.}~\bibnamefont {Reehorst}}, \bibinfo {author} {\bibfnamefont {S.}~\bibnamefont {Rychkov}}, \bibinfo {author} {\bibfnamefont {D.}~\bibnamefont {Simmons-Duffin}}, \bibinfo {author} {\bibfnamefont {B.}~\bibnamefont {Sirois}}, \bibinfo {author} {\bibfnamefont {N.}~\bibnamefont {Su}}, \ and\ \bibinfo {author} {\bibfnamefont {B.}~\bibnamefont {van Rees}},\ }\href {\doibase 10.21468/SciPostPhys.11.3.072} {\bibfield  {journal} {\bibinfo  {journal} {SciPost Phys.}\ }\textbf {\bibinfo {volume} {11}},\ \bibinfo {pages} {072} (\bibinfo {year} {2021})},\ \Eprint {http://arxiv.org/abs/2104.09518} {arXiv:2104.09518 [hep-th]} \BibitemShut {NoStop}%
\bibitem [{\citenamefont {Xu}(2013)}]{Xu2013AnAP}%
  \BibitemOpen
  \bibfield  {author} {\bibinfo {author} {\bibfnamefont {G.}~\bibnamefont {Xu}},\ }\href {https://api.semanticscholar.org/CorpusID:1283967} {\bibfield  {journal} {\bibinfo  {journal} {Appl. Math. Comput.}\ }\textbf {\bibinfo {volume} {219}},\ \bibinfo {pages} {4560} (\bibinfo {year} {2013})}\BibitemShut {NoStop}%
\bibitem [{\citenamefont {Protopopescu}\ \emph {et~al.}(1973)\citenamefont {Protopopescu}, \citenamefont {Alston-Garnjost}, \citenamefont {Barbaro-Galtieri}, \citenamefont {Flatte}, \citenamefont {Friedman}, \citenamefont {Lasinski}, \citenamefont {Lynch}, \citenamefont {Rabin},\ and\ \citenamefont {Solmitz}}]{Protopopescu:1973sh}%
  \BibitemOpen
  \bibfield  {author} {\bibinfo {author} {\bibfnamefont {S.}~\bibnamefont {Protopopescu}}, \bibinfo {author} {\bibfnamefont {M.}~\bibnamefont {Alston-Garnjost}}, \bibinfo {author} {\bibfnamefont {A.}~\bibnamefont {Barbaro-Galtieri}}, \bibinfo {author} {\bibfnamefont {S.~M.}\ \bibnamefont {Flatte}}, \bibinfo {author} {\bibfnamefont {J.}~\bibnamefont {Friedman}}, \bibinfo {author} {\bibfnamefont {T.}~\bibnamefont {Lasinski}}, \bibinfo {author} {\bibfnamefont {G.}~\bibnamefont {Lynch}}, \bibinfo {author} {\bibfnamefont {M.}~\bibnamefont {Rabin}}, \ and\ \bibinfo {author} {\bibfnamefont {F.}~\bibnamefont {Solmitz}},\ }\href {\doibase 10.1103/PhysRevD.7.1279} {\bibfield  {journal} {\bibinfo  {journal} {Phys. Rev. D}\ }\textbf {\bibinfo {volume} {7}},\ \bibinfo {pages} {1279} (\bibinfo {year} {1973})}\BibitemShut {NoStop}%
\bibitem [{\citenamefont {Estabrooks}\ and\ \citenamefont {Martin}(1974)}]{Estabrooks:1974vu}%
  \BibitemOpen
  \bibfield  {author} {\bibinfo {author} {\bibfnamefont {P.}~\bibnamefont {Estabrooks}}\ and\ \bibinfo {author} {\bibfnamefont {A.~D.}\ \bibnamefont {Martin}},\ }\href {\doibase 10.1016/0550-3213(74)90488-X} {\bibfield  {journal} {\bibinfo  {journal} {Nucl. Phys. B}\ }\textbf {\bibinfo {volume} {79}},\ \bibinfo {pages} {301} (\bibinfo {year} {1974})}\BibitemShut {NoStop}%
\bibitem [{\citenamefont {Grayer}\ \emph {et~al.}(1974)\citenamefont {Grayer} \emph {et~al.}}]{Grayer:1974cr}%
  \BibitemOpen
  \bibfield  {author} {\bibinfo {author} {\bibfnamefont {G.}~\bibnamefont {Grayer}} \emph {et~al.},\ }\href {\doibase 10.1016/0550-3213(74)90545-8} {\bibfield  {journal} {\bibinfo  {journal} {Nucl. Phys. B}\ }\textbf {\bibinfo {volume} {75}},\ \bibinfo {pages} {189} (\bibinfo {year} {1974})}\BibitemShut {NoStop}%
\bibitem [{\citenamefont {Losty}\ \emph {et~al.}(1974)\citenamefont {Losty}, \citenamefont {Chaloupka}, \citenamefont {Ferrando}, \citenamefont {Montanet}, \citenamefont {Paul}, \citenamefont {Yaffe}, \citenamefont {Zieminski}, \citenamefont {Alitti}, \citenamefont {Gandois},\ and\ \citenamefont {Louie}}]{Losty:1973et}%
  \BibitemOpen
  \bibfield  {author} {\bibinfo {author} {\bibfnamefont {M.}~\bibnamefont {Losty}}, \bibinfo {author} {\bibfnamefont {V.}~\bibnamefont {Chaloupka}}, \bibinfo {author} {\bibfnamefont {A.}~\bibnamefont {Ferrando}}, \bibinfo {author} {\bibfnamefont {L.}~\bibnamefont {Montanet}}, \bibinfo {author} {\bibfnamefont {E.}~\bibnamefont {Paul}}, \bibinfo {author} {\bibfnamefont {D.}~\bibnamefont {Yaffe}}, \bibinfo {author} {\bibfnamefont {A.}~\bibnamefont {Zieminski}}, \bibinfo {author} {\bibfnamefont {J.}~\bibnamefont {Alitti}}, \bibinfo {author} {\bibfnamefont {B.}~\bibnamefont {Gandois}}, \ and\ \bibinfo {author} {\bibfnamefont {J.}~\bibnamefont {Louie}},\ }\href {\doibase 10.1016/0550-3213(74)90131-X} {\bibfield  {journal} {\bibinfo  {journal} {Nucl. Phys. B}\ }\textbf {\bibinfo {volume} {69}},\ \bibinfo {pages} {185} (\bibinfo {year} {1974})}\BibitemShut {NoStop}%
\bibitem [{\citenamefont {Hoogland}\ \emph {et~al.}(1977)\citenamefont {Hoogland} \emph {et~al.}}]{Hoogland:1977kt}%
  \BibitemOpen
  \bibfield  {author} {\bibinfo {author} {\bibfnamefont {W.}~\bibnamefont {Hoogland}} \emph {et~al.},\ }\href {\doibase 10.1016/0550-3213(77)90154-7} {\bibfield  {journal} {\bibinfo  {journal} {Nucl. Phys. B}\ }\textbf {\bibinfo {volume} {126}},\ \bibinfo {pages} {109} (\bibinfo {year} {1977})}\BibitemShut {NoStop}%
\bibitem [{\citenamefont {Pelaez}\ and\ \citenamefont {Yndurain}(2005)}]{Pelaez:2004vs}%
  \BibitemOpen
  \bibfield  {author} {\bibinfo {author} {\bibfnamefont {J.~R.}\ \bibnamefont {Pelaez}}\ and\ \bibinfo {author} {\bibfnamefont {F.~J.}\ \bibnamefont {Yndurain}},\ }\href {\doibase 10.1103/PhysRevD.71.074016} {\bibfield  {journal} {\bibinfo  {journal} {Phys. Rev. D}\ }\textbf {\bibinfo {volume} {71}},\ \bibinfo {pages} {074016} (\bibinfo {year} {2005})},\ \Eprint {http://arxiv.org/abs/hep-ph/0411334} {arXiv:hep-ph/0411334} \BibitemShut {NoStop}%
\bibitem [{\citenamefont {Batley}\ \emph {et~al.}(2010)\citenamefont {Batley} \emph {et~al.}}]{Batley:2010zza}%
  \BibitemOpen
  \bibfield  {author} {\bibinfo {author} {\bibfnamefont {J.}~\bibnamefont {Batley}} \emph {et~al.} (\bibinfo {collaboration} {NA48/2}),\ }\href {\doibase 10.1140/epjc/s10052-010-1480-6} {\bibfield  {journal} {\bibinfo  {journal} {Eur. Phys. J. C}\ }\textbf {\bibinfo {volume} {70}},\ \bibinfo {pages} {635} (\bibinfo {year} {2010})}\BibitemShut {NoStop}%
\bibitem [{\citenamefont {Blum}\ \emph {et~al.}(2021)\citenamefont {Blum} \emph {et~al.}}]{RBC:2021acc}%
  \BibitemOpen
  \bibfield  {author} {\bibinfo {author} {\bibfnamefont {T.}~\bibnamefont {Blum}} \emph {et~al.} (\bibinfo {collaboration} {RBC, UKQCD}),\ }\href {\doibase 10.1103/PhysRevD.104.114506} {\bibfield  {journal} {\bibinfo  {journal} {Phys. Rev. D}\ }\textbf {\bibinfo {volume} {104}},\ \bibinfo {pages} {114506} (\bibinfo {year} {2021})},\ \Eprint {http://arxiv.org/abs/2103.15131} {arXiv:2103.15131 [hep-lat]} \BibitemShut {NoStop}%
\bibitem [{\citenamefont {de~Troconiz}\ and\ \citenamefont {Yndurain}(2005)}]{deTroconiz:2004yzs}%
  \BibitemOpen
  \bibfield  {author} {\bibinfo {author} {\bibfnamefont {J.~F.}\ \bibnamefont {de~Troconiz}}\ and\ \bibinfo {author} {\bibfnamefont {F.~J.}\ \bibnamefont {Yndurain}},\ }\href {\doibase 10.1103/PhysRevD.71.073008} {\bibfield  {journal} {\bibinfo  {journal} {Phys. Rev. D}\ }\textbf {\bibinfo {volume} {71}},\ \bibinfo {pages} {073008} (\bibinfo {year} {2005})},\ \Eprint {http://arxiv.org/abs/hep-ph/0402285} {arXiv:hep-ph/0402285} \BibitemShut {NoStop}%
\bibitem [{\citenamefont {Colangelo}\ \emph {et~al.}(2019)\citenamefont {Colangelo}, \citenamefont {Hoferichter},\ and\ \citenamefont {Stoffer}}]{Colangelo:2018mtw}%
  \BibitemOpen
  \bibfield  {author} {\bibinfo {author} {\bibfnamefont {G.}~\bibnamefont {Colangelo}}, \bibinfo {author} {\bibfnamefont {M.}~\bibnamefont {Hoferichter}}, \ and\ \bibinfo {author} {\bibfnamefont {P.}~\bibnamefont {Stoffer}},\ }\href {\doibase 10.1007/JHEP02(2019)006} {\bibfield  {journal} {\bibinfo  {journal} {JHEP}\ }\textbf {\bibinfo {volume} {02}},\ \bibinfo {pages} {006} (\bibinfo {year} {2019})},\ \Eprint {http://arxiv.org/abs/1810.00007} {arXiv:1810.00007 [hep-ph]} \BibitemShut {NoStop}%
\bibitem [{\citenamefont {Boyle}\ \emph {et~al.}(2024)\citenamefont {Boyle}, \citenamefont {Erben}, \citenamefont {G\"ulpers}, \citenamefont {Hansen}, \citenamefont {Joswig}, \citenamefont {Lachini}, \citenamefont {Marshall},\ and\ \citenamefont {Portelli}}]{Boyle:2024grr}%
  \BibitemOpen
  \bibfield  {author} {\bibinfo {author} {\bibfnamefont {P.}~\bibnamefont {Boyle}}, \bibinfo {author} {\bibfnamefont {F.}~\bibnamefont {Erben}}, \bibinfo {author} {\bibfnamefont {V.}~\bibnamefont {G\"ulpers}}, \bibinfo {author} {\bibfnamefont {M.~T.}\ \bibnamefont {Hansen}}, \bibinfo {author} {\bibfnamefont {F.}~\bibnamefont {Joswig}}, \bibinfo {author} {\bibfnamefont {N.~P.}\ \bibnamefont {Lachini}}, \bibinfo {author} {\bibfnamefont {M.}~\bibnamefont {Marshall}}, \ and\ \bibinfo {author} {\bibfnamefont {A.}~\bibnamefont {Portelli}},\ }\href@noop {} {\  (\bibinfo {year} {2024})},\ \Eprint {http://arxiv.org/abs/2406.19193} {arXiv:2406.19193 [hep-lat]} \BibitemShut {NoStop}%
\bibitem [{\citenamefont {Bijnens}\ \emph {et~al.}(1996)\citenamefont {Bijnens}, \citenamefont {Colangelo}, \citenamefont {Ecker}, \citenamefont {Gasser},\ and\ \citenamefont {Sainio}}]{Bijnens:1995yn}%
  \BibitemOpen
  \bibfield  {author} {\bibinfo {author} {\bibfnamefont {J.}~\bibnamefont {Bijnens}}, \bibinfo {author} {\bibfnamefont {G.}~\bibnamefont {Colangelo}}, \bibinfo {author} {\bibfnamefont {G.}~\bibnamefont {Ecker}}, \bibinfo {author} {\bibfnamefont {J.}~\bibnamefont {Gasser}}, \ and\ \bibinfo {author} {\bibfnamefont {M.}~\bibnamefont {Sainio}},\ }\href {\doibase 10.1016/0370-2693(96)00165-7} {\bibfield  {journal} {\bibinfo  {journal} {Phys. Lett. B}\ }\textbf {\bibinfo {volume} {374}},\ \bibinfo {pages} {210} (\bibinfo {year} {1996})},\ \Eprint {http://arxiv.org/abs/hep-ph/9511397} {arXiv:hep-ph/9511397} \BibitemShut {NoStop}%
\bibitem [{\citenamefont {Colangelo}\ \emph {et~al.}(2001)\citenamefont {Colangelo}, \citenamefont {Gasser},\ and\ \citenamefont {Leutwyler}}]{Colangelo:2001df}%
  \BibitemOpen
  \bibfield  {author} {\bibinfo {author} {\bibfnamefont {G.}~\bibnamefont {Colangelo}}, \bibinfo {author} {\bibfnamefont {J.}~\bibnamefont {Gasser}}, \ and\ \bibinfo {author} {\bibfnamefont {H.}~\bibnamefont {Leutwyler}},\ }\href {\doibase 10.1016/S0550-3213(01)00147-X} {\bibfield  {journal} {\bibinfo  {journal} {Nucl. Phys. B}\ }\textbf {\bibinfo {volume} {603}},\ \bibinfo {pages} {125} (\bibinfo {year} {2001})},\ \Eprint {http://arxiv.org/abs/hep-ph/0103088} {arXiv:hep-ph/0103088} \BibitemShut {NoStop}%
\bibitem [{\citenamefont {Aaij}\ \emph {et~al.}(2020)\citenamefont {Aaij} \emph {et~al.}}]{LHCb:2019sus}%
  \BibitemOpen
  \bibfield  {author} {\bibinfo {author} {\bibfnamefont {R.}~\bibnamefont {Aaij}} \emph {et~al.} (\bibinfo {collaboration} {LHCb}),\ }\href {\doibase 10.1103/PhysRevD.101.012006} {\bibfield  {journal} {\bibinfo  {journal} {Phys. Rev. D}\ }\textbf {\bibinfo {volume} {101}},\ \bibinfo {pages} {012006} (\bibinfo {year} {2020})},\ \Eprint {http://arxiv.org/abs/1909.05212} {arXiv:1909.05212 [hep-ex]} \BibitemShut {NoStop}%
\bibitem [{\citenamefont {Elias~Mir\'o}\ \emph {et~al.}(2019{\natexlab{a}})\citenamefont {Elias~Mir\'o}, \citenamefont {Guerrieri}, \citenamefont {Hebbar}, \citenamefont {Penedones},\ and\ \citenamefont {Vieira}}]{FluxTube}%
  \BibitemOpen
  \bibfield  {author} {\bibinfo {author} {\bibfnamefont {J.}~\bibnamefont {Elias~Mir\'o}}, \bibinfo {author} {\bibfnamefont {A.~L.}\ \bibnamefont {Guerrieri}}, \bibinfo {author} {\bibfnamefont {A.}~\bibnamefont {Hebbar}}, \bibinfo {author} {\bibfnamefont {J.}~\bibnamefont {Penedones}}, \ and\ \bibinfo {author} {\bibfnamefont {P.}~\bibnamefont {Vieira}},\ }\href {\doibase 10.1103/PhysRevLett.123.221602} {\bibfield  {journal} {\bibinfo  {journal} {Phys. Rev. Lett.}\ }\textbf {\bibinfo {volume} {123}},\ \bibinfo {pages} {221602} (\bibinfo {year} {2019}{\natexlab{a}})},\ \Eprint {http://arxiv.org/abs/1906.08098} {arXiv:1906.08098 [hep-th]} \BibitemShut {NoStop}%
\bibitem [{\citenamefont {Biswas}\ \emph {et~al.}(1967)\citenamefont {Biswas}, \citenamefont {Cason}, \citenamefont {Derado}, \citenamefont {Kenney}, \citenamefont {Poirier},\ and\ \citenamefont {Shephard}}]{Biswas:1967mpl}%
  \BibitemOpen
  \bibfield  {author} {\bibinfo {author} {\bibfnamefont {N.~N.}\ \bibnamefont {Biswas}}, \bibinfo {author} {\bibfnamefont {N.~M.}\ \bibnamefont {Cason}}, \bibinfo {author} {\bibfnamefont {I.}~\bibnamefont {Derado}}, \bibinfo {author} {\bibfnamefont {V.~P.}\ \bibnamefont {Kenney}}, \bibinfo {author} {\bibfnamefont {J.~A.}\ \bibnamefont {Poirier}}, \ and\ \bibinfo {author} {\bibfnamefont {W.~D.}\ \bibnamefont {Shephard}},\ }\href {\doibase 10.1103/PhysRevLett.18.273} {\bibfield  {journal} {\bibinfo  {journal} {Phys. Rev. Lett.}\ }\textbf {\bibinfo {volume} {18}},\ \bibinfo {pages} {273} (\bibinfo {year} {1967})}\BibitemShut {NoStop}%
\bibitem [{\citenamefont {Robertson}\ \emph {et~al.}(1973)\citenamefont {Robertson}, \citenamefont {Walker},\ and\ \citenamefont {Davis}}]{Robertson:1973tk}%
  \BibitemOpen
  \bibfield  {author} {\bibinfo {author} {\bibfnamefont {W.~J.}\ \bibnamefont {Robertson}}, \bibinfo {author} {\bibfnamefont {W.~D.}\ \bibnamefont {Walker}}, \ and\ \bibinfo {author} {\bibfnamefont {J.~L.}\ \bibnamefont {Davis}},\ }\href {\doibase 10.1103/PhysRevD.7.2554} {\bibfield  {journal} {\bibinfo  {journal} {Phys. Rev. D}\ }\textbf {\bibinfo {volume} {7}},\ \bibinfo {pages} {2554} (\bibinfo {year} {1973})}\BibitemShut {NoStop}%
\bibitem [{\citenamefont {Cohen}\ \emph {et~al.}(1973)\citenamefont {Cohen}, \citenamefont {Ferbel}, \citenamefont {Slattery},\ and\ \citenamefont {Werner}}]{Cohen:1973yx}%
  \BibitemOpen
  \bibfield  {author} {\bibinfo {author} {\bibfnamefont {D.~H.}\ \bibnamefont {Cohen}}, \bibinfo {author} {\bibfnamefont {T.}~\bibnamefont {Ferbel}}, \bibinfo {author} {\bibfnamefont {P.}~\bibnamefont {Slattery}}, \ and\ \bibinfo {author} {\bibfnamefont {B.}~\bibnamefont {Werner}},\ }\href {\doibase 10.1103/PhysRevD.7.661} {\bibfield  {journal} {\bibinfo  {journal} {Phys. Rev. D}\ }\textbf {\bibinfo {volume} {7}},\ \bibinfo {pages} {661} (\bibinfo {year} {1973})}\BibitemShut {NoStop}%
\bibitem [{\citenamefont {Hanlon}\ \emph {et~al.}(1976)\citenamefont {Hanlon} \emph {et~al.}}]{Hanlon:1976ct}%
  \BibitemOpen
  \bibfield  {author} {\bibinfo {author} {\bibfnamefont {J.}~\bibnamefont {Hanlon}} \emph {et~al.},\ }\href {\doibase 10.1103/PhysRevLett.37.967} {\bibfield  {journal} {\bibinfo  {journal} {Phys. Rev. Lett.}\ }\textbf {\bibinfo {volume} {37}},\ \bibinfo {pages} {967} (\bibinfo {year} {1976})}\BibitemShut {NoStop}%
\bibitem [{\citenamefont {Abramowicz}\ \emph {et~al.}(1980)\citenamefont {Abramowicz} \emph {et~al.}}]{Abramowicz:1979ca}%
  \BibitemOpen
  \bibfield  {author} {\bibinfo {author} {\bibfnamefont {H.}~\bibnamefont {Abramowicz}} \emph {et~al.},\ }\href {\doibase 10.1016/0550-3213(80)90489-7} {\bibfield  {journal} {\bibinfo  {journal} {Nucl. Phys. B}\ }\textbf {\bibinfo {volume} {166}},\ \bibinfo {pages} {62} (\bibinfo {year} {1980})}\BibitemShut {NoStop}%
\bibitem [{\citenamefont {Zakharov}\ and\ \citenamefont {Sergeev}(1984)}]{Zakharov:1984nq}%
  \BibitemOpen
  \bibfield  {author} {\bibinfo {author} {\bibfnamefont {B.~G.}\ \bibnamefont {Zakharov}}\ and\ \bibinfo {author} {\bibfnamefont {V.~N.}\ \bibnamefont {Sergeev}},\ }\href@noop {} {\bibfield  {journal} {\bibinfo  {journal} {Yad. Fiz.}\ }\textbf {\bibinfo {volume} {39}},\ \bibinfo {pages} {707} (\bibinfo {year} {1984})}\BibitemShut {NoStop}%
\bibitem [{\citenamefont {Bhat}\ \emph {et~al.}(2024)\citenamefont {Bhat}, \citenamefont {Chowdhury}, \citenamefont {Sinha}, \citenamefont {Tiwari},\ and\ \citenamefont {Zahed}}]{Bhat:2023puy}%
  \BibitemOpen
  \bibfield  {author} {\bibinfo {author} {\bibfnamefont {F.}~\bibnamefont {Bhat}}, \bibinfo {author} {\bibfnamefont {D.}~\bibnamefont {Chowdhury}}, \bibinfo {author} {\bibfnamefont {A.}~\bibnamefont {Sinha}}, \bibinfo {author} {\bibfnamefont {S.}~\bibnamefont {Tiwari}}, \ and\ \bibinfo {author} {\bibfnamefont {A.}~\bibnamefont {Zahed}},\ }\href {\doibase 10.1007/JHEP03(2024)157} {\bibfield  {journal} {\bibinfo  {journal} {JHEP}\ }\textbf {\bibinfo {volume} {03}},\ \bibinfo {pages} {157} (\bibinfo {year} {2024})},\ \Eprint {http://arxiv.org/abs/2311.03451} {arXiv:2311.03451 [hep-th]} \BibitemShut {NoStop}%
\bibitem [{\citenamefont {Gumus}(2023)}]{Gumus:2023xbs}%
  \BibitemOpen
  \bibfield  {author} {\bibinfo {author} {\bibfnamefont {M.~A.}\ \bibnamefont {Gumus}},\ }\emph {\bibinfo {title} {{Mapping out EFTs with analytic S-matrix}}},\ \href@noop {} {Ph.D. thesis},\ \bibinfo  {school} {SISSA, Italy, SISSA, SISSA, Trieste} (\bibinfo {year} {2023})\BibitemShut {NoStop}%
\bibitem [{\citenamefont {Elias~Mir\'o}\ \emph {et~al.}()\citenamefont {Elias~Mir\'o}, \citenamefont {Guerrieri},\ and\ \citenamefont {Gumus}}]{Reggeworkinprogress}%
  \BibitemOpen
  \bibfield  {author} {\bibinfo {author} {\bibfnamefont {J.}~\bibnamefont {Elias~Mir\'o}}, \bibinfo {author} {\bibfnamefont {A.}~\bibnamefont {Guerrieri}}, \ and\ \bibinfo {author} {\bibfnamefont {M.}~\bibnamefont {Gumus}},\ }\href@noop {} {\bibinfo  {journal} {{work in progress}}\ }\BibitemShut {NoStop}%
\bibitem [{\citenamefont {Correia}\ \emph {et~al.}()\citenamefont {Correia}, \citenamefont {Georgoudis},\ and\ \citenamefont {Guerrieri}}]{froissartbound}%
  \BibitemOpen
\bibfield  {journal} {  }\bibfield  {author} {\bibinfo {author} {\bibfnamefont {M.}~\bibnamefont {Correia}}, \bibinfo {author} {\bibfnamefont {A.}~\bibnamefont {Georgoudis}}, \ and\ \bibinfo {author} {\bibfnamefont {A.}~\bibnamefont {Guerrieri}},\ }\href@noop {} {\bibinfo  {journal} {{work in progress}}\ }\BibitemShut {NoStop}%
\bibitem [{\citenamefont {Caprini}\ \emph {et~al.}(2012)\citenamefont {Caprini}, \citenamefont {Colangelo},\ and\ \citenamefont {Leutwyler}}]{Caprini:2011ky}%
  \BibitemOpen
\bibfield  {journal} {  }\bibfield  {author} {\bibinfo {author} {\bibfnamefont {I.}~\bibnamefont {Caprini}}, \bibinfo {author} {\bibfnamefont {G.}~\bibnamefont {Colangelo}}, \ and\ \bibinfo {author} {\bibfnamefont {H.}~\bibnamefont {Leutwyler}},\ }\href {\doibase 10.1140/epjc/s10052-012-1860-1} {\bibfield  {journal} {\bibinfo  {journal} {Eur. Phys. J. C}\ }\textbf {\bibinfo {volume} {72}},\ \bibinfo {pages} {1860} (\bibinfo {year} {2012})},\ \Eprint {http://arxiv.org/abs/1111.7160} {arXiv:1111.7160 [hep-ph]} \BibitemShut {NoStop}%
\bibitem [{\citenamefont {Caprini}\ \emph {et~al.}(2006)\citenamefont {Caprini}, \citenamefont {Colangelo},\ and\ \citenamefont {Leutwyler}}]{Caprini:2005zr}%
  \BibitemOpen
  \bibfield  {author} {\bibinfo {author} {\bibfnamefont {I.}~\bibnamefont {Caprini}}, \bibinfo {author} {\bibfnamefont {G.}~\bibnamefont {Colangelo}}, \ and\ \bibinfo {author} {\bibfnamefont {H.}~\bibnamefont {Leutwyler}},\ }\href {\doibase 10.1103/PhysRevLett.96.132001} {\bibfield  {journal} {\bibinfo  {journal} {Phys. Rev. Lett.}\ }\textbf {\bibinfo {volume} {96}},\ \bibinfo {pages} {132001} (\bibinfo {year} {2006})},\ \Eprint {http://arxiv.org/abs/hep-ph/0512364} {arXiv:hep-ph/0512364} \BibitemShut {NoStop}%
\bibitem [{\citenamefont {Guerrieri}\ \emph {et~al.}()\citenamefont {Guerrieri}, \citenamefont {Kelian},\ and\ \citenamefont {Ning}}]{fit2_work_in_progress}%
  \BibitemOpen
  \bibfield  {author} {\bibinfo {author} {\bibfnamefont {A.}~\bibnamefont {Guerrieri}}, \bibinfo {author} {\bibfnamefont {H.}~\bibnamefont {Kelian}}, \ and\ \bibinfo {author} {\bibfnamefont {S.}~\bibnamefont {Ning}},\ }\href@noop {} {\bibinfo  {journal} {{work in progress}}\ }\BibitemShut {NoStop}%
\bibitem [{\citenamefont {Elias~Mir\'o}\ \emph {et~al.}(2019{\natexlab{b}})\citenamefont {Elias~Mir\'o}, \citenamefont {Guerrieri}, \citenamefont {Hebbar}, \citenamefont {Penedones},\ and\ \citenamefont {Vieira}}]{EliasMiro:2019kyf}%
  \BibitemOpen
\bibfield  {journal} {  }\bibfield  {author} {\bibinfo {author} {\bibfnamefont {J.}~\bibnamefont {Elias~Mir\'o}}, \bibinfo {author} {\bibfnamefont {A.~L.}\ \bibnamefont {Guerrieri}}, \bibinfo {author} {\bibfnamefont {A.}~\bibnamefont {Hebbar}}, \bibinfo {author} {\bibfnamefont {J.~a.}\ \bibnamefont {Penedones}}, \ and\ \bibinfo {author} {\bibfnamefont {P.}~\bibnamefont {Vieira}},\ }\href {\doibase 10.1103/PhysRevLett.123.221602} {\bibfield  {journal} {\bibinfo  {journal} {Phys. Rev. Lett.}\ }\textbf {\bibinfo {volume} {123}},\ \bibinfo {pages} {221602} (\bibinfo {year} {2019}{\natexlab{b}})},\ \Eprint {http://arxiv.org/abs/1906.08098} {arXiv:1906.08098 [hep-th]} \BibitemShut {NoStop}%
\bibitem [{\citenamefont {Guerrieri}\ \emph {et~al.}(2021{\natexlab{b}})\citenamefont {Guerrieri}, \citenamefont {Penedones},\ and\ \citenamefont {Vieira}}]{Guerrieri:2021ivu}%
  \BibitemOpen
  \bibfield  {author} {\bibinfo {author} {\bibfnamefont {A.}~\bibnamefont {Guerrieri}}, \bibinfo {author} {\bibfnamefont {J.}~\bibnamefont {Penedones}}, \ and\ \bibinfo {author} {\bibfnamefont {P.}~\bibnamefont {Vieira}},\ }\href {\doibase 10.1103/PhysRevLett.127.081601} {\bibfield  {journal} {\bibinfo  {journal} {Phys. Rev. Lett.}\ }\textbf {\bibinfo {volume} {127}},\ \bibinfo {pages} {081601} (\bibinfo {year} {2021}{\natexlab{b}})},\ \Eprint {http://arxiv.org/abs/2102.02847} {arXiv:2102.02847 [hep-th]} \BibitemShut {NoStop}%
\bibitem [{\citenamefont {Guerrieri}\ \emph {et~al.}(2023{\natexlab{b}})\citenamefont {Guerrieri}, \citenamefont {Murali}, \citenamefont {Penedones},\ and\ \citenamefont {Vieira}}]{Guerrieri:2022sod}%
  \BibitemOpen
  \bibfield  {author} {\bibinfo {author} {\bibfnamefont {A.}~\bibnamefont {Guerrieri}}, \bibinfo {author} {\bibfnamefont {H.}~\bibnamefont {Murali}}, \bibinfo {author} {\bibfnamefont {J.}~\bibnamefont {Penedones}}, \ and\ \bibinfo {author} {\bibfnamefont {P.}~\bibnamefont {Vieira}},\ }\href {\doibase 10.1007/JHEP06(2023)064} {\bibfield  {journal} {\bibinfo  {journal} {JHEP}\ }\textbf {\bibinfo {volume} {06}},\ \bibinfo {pages} {064} (\bibinfo {year} {2023}{\natexlab{b}})},\ \Eprint {http://arxiv.org/abs/2212.00151} {arXiv:2212.00151 [hep-th]} \BibitemShut {NoStop}%
\bibitem [{\citenamefont {H\"aring}\ \emph {et~al.}(2024)\citenamefont {H\"aring}, \citenamefont {Hebbar}, \citenamefont {Karateev}, \citenamefont {Meineri},\ and\ \citenamefont {Penedones}}]{Haring:2022sdp}%
  \BibitemOpen
  \bibfield  {author} {\bibinfo {author} {\bibfnamefont {K.}~\bibnamefont {H\"aring}}, \bibinfo {author} {\bibfnamefont {A.}~\bibnamefont {Hebbar}}, \bibinfo {author} {\bibfnamefont {D.}~\bibnamefont {Karateev}}, \bibinfo {author} {\bibfnamefont {M.}~\bibnamefont {Meineri}}, \ and\ \bibinfo {author} {\bibfnamefont {J.~a.}\ \bibnamefont {Penedones}},\ }\href {\doibase 10.1007/JHEP10(2024)103} {\bibfield  {journal} {\bibinfo  {journal} {JHEP}\ }\textbf {\bibinfo {volume} {2410}},\ \bibinfo {pages} {103} (\bibinfo {year} {2024})},\ \Eprint {http://arxiv.org/abs/2211.05795} {arXiv:2211.05795 [hep-th]} \BibitemShut {NoStop}%
\bibitem [{\citenamefont {Acanfora}\ \emph {et~al.}(2024)\citenamefont {Acanfora}, \citenamefont {Guerrieri}, \citenamefont {H\"aring},\ and\ \citenamefont {Karateev}}]{Acanfora:2023axz}%
  \BibitemOpen
  \bibfield  {author} {\bibinfo {author} {\bibfnamefont {F.}~\bibnamefont {Acanfora}}, \bibinfo {author} {\bibfnamefont {A.}~\bibnamefont {Guerrieri}}, \bibinfo {author} {\bibfnamefont {K.}~\bibnamefont {H\"aring}}, \ and\ \bibinfo {author} {\bibfnamefont {D.}~\bibnamefont {Karateev}},\ }\href {\doibase 10.1007/JHEP03(2024)028} {\bibfield  {journal} {\bibinfo  {journal} {JHEP}\ }\textbf {\bibinfo {volume} {03}},\ \bibinfo {pages} {028} (\bibinfo {year} {2024})},\ \Eprint {http://arxiv.org/abs/2310.06027} {arXiv:2310.06027 [hep-th]} \BibitemShut {NoStop}%
\bibitem [{\citenamefont {Karateev}\ \emph {et~al.}(2020)\citenamefont {Karateev}, \citenamefont {Kuhn},\ and\ \citenamefont {Penedones}}]{Karateev:2019ymz}%
  \BibitemOpen
  \bibfield  {author} {\bibinfo {author} {\bibfnamefont {D.}~\bibnamefont {Karateev}}, \bibinfo {author} {\bibfnamefont {S.}~\bibnamefont {Kuhn}}, \ and\ \bibinfo {author} {\bibfnamefont {J.~a.}\ \bibnamefont {Penedones}},\ }\href {\doibase 10.1007/JHEP07(2020)035} {\bibfield  {journal} {\bibinfo  {journal} {JHEP}\ }\textbf {\bibinfo {volume} {07}},\ \bibinfo {pages} {035} (\bibinfo {year} {2020})},\ \Eprint {http://arxiv.org/abs/1912.08940} {arXiv:1912.08940 [hep-th]} \BibitemShut {NoStop}%
\bibitem [{\citenamefont {Correia}\ \emph {et~al.}(2023)\citenamefont {Correia}, \citenamefont {Penedones},\ and\ \citenamefont {Vuignier}}]{Correia:2022dyp}%
  \BibitemOpen
  \bibfield  {author} {\bibinfo {author} {\bibfnamefont {M.}~\bibnamefont {Correia}}, \bibinfo {author} {\bibfnamefont {J.}~\bibnamefont {Penedones}}, \ and\ \bibinfo {author} {\bibfnamefont {A.}~\bibnamefont {Vuignier}},\ }\href {\doibase 10.1007/JHEP08(2023)108} {\bibfield  {journal} {\bibinfo  {journal} {JHEP}\ }\textbf {\bibinfo {volume} {08}},\ \bibinfo {pages} {108} (\bibinfo {year} {2023})},\ \Eprint {http://arxiv.org/abs/2212.03917} {arXiv:2212.03917 [hep-th]} \BibitemShut {NoStop}%
\bibitem [{\citenamefont {Cordova}\ \emph {et~al.}(2024)\citenamefont {Cordova}, \citenamefont {Correia}, \citenamefont {Georgoudis},\ and\ \citenamefont {Vuignier}}]{Cordova:2023wjp}%
  \BibitemOpen
  \bibfield  {author} {\bibinfo {author} {\bibfnamefont {L.}~\bibnamefont {Cordova}}, \bibinfo {author} {\bibfnamefont {M.}~\bibnamefont {Correia}}, \bibinfo {author} {\bibfnamefont {A.}~\bibnamefont {Georgoudis}}, \ and\ \bibinfo {author} {\bibfnamefont {A.}~\bibnamefont {Vuignier}},\ }\href {\doibase 10.1007/JHEP01(2024)093} {\bibfield  {journal} {\bibinfo  {journal} {JHEP}\ }\textbf {\bibinfo {volume} {01}},\ \bibinfo {pages} {093} (\bibinfo {year} {2024})},\ \Eprint {http://arxiv.org/abs/2311.03031} {arXiv:2311.03031 [hep-th]} \BibitemShut {NoStop}%
\bibitem [{\citenamefont {Chen}\ \emph {et~al.}(2022)\citenamefont {Chen}, \citenamefont {Fitzpatrick},\ and\ \citenamefont {Karateev}}]{Chen:2021pgx}%
  \BibitemOpen
  \bibfield  {author} {\bibinfo {author} {\bibfnamefont {H.}~\bibnamefont {Chen}}, \bibinfo {author} {\bibfnamefont {A.~L.}\ \bibnamefont {Fitzpatrick}}, \ and\ \bibinfo {author} {\bibfnamefont {D.}~\bibnamefont {Karateev}},\ }\href {\doibase 10.1007/JHEP02(2022)146} {\bibfield  {journal} {\bibinfo  {journal} {JHEP}\ }\textbf {\bibinfo {volume} {02}},\ \bibinfo {pages} {146} (\bibinfo {year} {2022})},\ \Eprint {http://arxiv.org/abs/2107.10286} {arXiv:2107.10286 [hep-th]} \BibitemShut {NoStop}%
\bibitem [{\citenamefont {Gabai}\ and\ \citenamefont {Yin}(2022)}]{Gabai:2019ryw}%
  \BibitemOpen
  \bibfield  {author} {\bibinfo {author} {\bibfnamefont {B.}~\bibnamefont {Gabai}}\ and\ \bibinfo {author} {\bibfnamefont {X.}~\bibnamefont {Yin}},\ }\href {\doibase 10.1007/JHEP10(2022)168} {\bibfield  {journal} {\bibinfo  {journal} {JHEP}\ }\textbf {\bibinfo {volume} {10}},\ \bibinfo {pages} {168} (\bibinfo {year} {2022})},\ \Eprint {http://arxiv.org/abs/1905.00710} {arXiv:1905.00710 [hep-th]} \BibitemShut {NoStop}%
\end{thebibliography}%

\end{document}